\documentclass[10pt,letterpaper]{article} 
\usepackage[english]{babel}
\newcommand{\cals}{\cal S}
\newcommand{\calm}{\cal M}
\newcommand{\lb}{\langle}
\newcommand{\rb}{\rangle}

\newcommand{\beq}{\begin{equation}}
\newcommand{\eeq}{\end{equation}}
\newcommand{\lbl}{\label}
\newcommand{\beqnar}{\begin{eqnarray}}
\newcommand{\eeqnar}{\end{eqnarray}}
\newcommand{\beqnars}{\begin{eqnarray*}}
\newcommand{\eeqnars}{\end{eqnarray*}}

\newcommand{\goesto}{\rightarrow}
\newcommand{\s}{\\[1ex]}
\newcommand{\re}[1]{(\ref{#1})}
\newcommand{\q}{\quad}

\newcommand{\tnr}{\otimes} 
\newcommand{\tr}{\mbox{tr }}

\newcommand{\trM}{\mbox{tr}_M\,}
\newcommand{\trS}{\mbox{tr}_S\,}

\newtheorem{definition}{Definition}

\newcommand{\Exp}{{\rm Exp}}
\newcommand{\rhat}{\hat{r}}
\newcommand{\bR}{{\bf R}} 
\begin{document} 
%
\begin{center}
\Large
\bf
What do quantum ``weak'' measurements actually measure? 
\\
\large
\rm 
\ 
\\
Stephen Parrott%
\footnote{For contact information, go to http://www.math.umb.edu/$\sim$sp}\\
\ 
\s
\normalsize 
September 30, 2009
\end{center}
\begin{abstract}
A precise definition of ``weak [quantum] measurements'' and 
``weak value'' (of a quantum observable) is offered, and simple finite 
dimensional examples are given showing that
weak values are not unique and therefore probably do not correspond to any
physical attribute of the system being ``weakly'' measured, contrary to 
impressions given by most of the literature on weak measurements.

A possible mathematical error in the seminal paper introducing ``weak values'' 
is explicitly identified.  
A mathematically rigorous argument obtains results similar to, 
and more general than, the main result of that paper  and concludes that even in 
the infinite-dimensional context of that paper, weak values 
are not unique.   This implies that the ``usual'' formula for weak values
is not universal, but can apply only to specific physical situations.  

The paper is written in a more pedagogical and informal style than is usual
in the research literature in the hope that it might serve as an introduction
to weak values.
\end{abstract}
\section{Introduction}

I have just spent several months pondering the implications of a 2 cm 
stack of recent papers concerning ``weak measurements" 
in quantum mechanics.  They include papers originally introducing this 
concept,  various rederivations and extensions, experimental 
``weak'' measurements, and their use to supposedly resolve  
``Hardy's paradox''.  

I have come to the conclusion that the concept of ``weak measurement'',
as presented in this literature, cannot withstand careful scrutiny.
That does not mean that there can be no truth or value in it, but that
generally accepted interpretations of its meaning are mistaken.
This note explains why. 

It will be written from a semi-historical viewpoint.  ``Semi'' because 
it seems clearest to first 
introduce the basic ideas in a simple, finite-dimensional context. 
Then we shall describe the more complicated infinite-dimensional situation
in which they were first introduced, resulting in the widely quoted 
formulas $\lb \psi_f , A \psi_{in} \rb / \lb \psi_f, \psi_{in}\rb$ 
or $\Re(\lb \psi_f , A \psi_{in} \rb / \lb \psi_f, \psi_{in}\rb)$ 
for the ``weak'' 
value of the observable $A$, given that the pre-measurement state was $\psi_i$, 
and that after measurement the state was ``post-selected'' to be $\psi_f$.
(All of these terms will be precisely defined later; they are included here
only to orient readers who may have some prior acquaintance with weak 
values.)  Then we shall observe that though the concept of ``weak
measurement'' may have some valid uses, we see no good reason to think that 
the results of most ``weak''
measurements  will be given 
given by one of these formulas.  
\section{Reader's guide to this paper}
	The aim of this paper is to share what I have learned in my study of 
	``weak values'' over the past few months.  It is written primarily for 
	physicists.  (I am a mathematician.)  
	It is hoped that it could serve as an introduction to 
	``weak values''.  In order to avoid possible misunderstandings, 
	it begins at a level more elementary than would typically be 
	assumed by a journal article.  
However, the reader is assumed to be thoroughly familiar with quantum 
mechanics, particularly in finite dimensions at the level of the book 
{\em Quantum Computation and Quantum Information} by Nielsen and Chuang
\cite{N/C}.  
Readers who are already 
	familiar with ``weak values'' can skim or skip the introductory 
	Sections 3 and 4.

The main ideas of the paper are presented in Section 5, ``A simple 
finite-dimensional model''.  The main new result is the observation
that weak values are not unique. I suspect that for many readers, this will
be all they care to learn about weak values.

The presentation of that section 
may seem quite different from most 
presentations in the literature.  It would be natural for a reader 
unfamiliar with weak values to wonder  if it is oversimplified, particularly
since the model is finite dimensional, whereas most presentations in the 
literature are set in a more complicated infinite dimensional context.  
Section 6 discussing the traditional approach is written for such readers. 
It obtains the usual formula
for a weak value in the traditional infinite dimensional context 
in what we hope is a mathematically rigorous way.  

So-called ``postselection'' is fundamental to the traditional approach.  
We have described it in a mathematically precise way but have given no
examples.  That is because despite the mathematical simplicity of the concept,
we do not know of any physically realistic examples which could be 
presented using a reasonable space and  without elaborate diagrams (which we
lack the facilities to prepare efficiently). 
The most enlightening examples we have seen are related
to ``Hardy's paradox''.  Readable descriptions of that can be found in  
\cite{RevHardy} and \cite{Yokota}.

Section 9 presents another simple proof that ``weak values'' are not unique.
Since it is closer to the methods common in the literature (such as the seminal
paper \cite{AAV} of Aharonov, Albert, and Vaidman) than the previous
exposition, it may be more congenial to those already familiar with weak
values.  As described there, it was discovered after the rest of the paper
 had already been written and typeset.  Were I writing it again, I might
try to find a way to put this new proof at center stage, but 
at this stage, it doesn't seem worth the effort.  The exposition as originally
written still seems more suitable for beginners. 

Appendices 1 and 2 discuss mathematical subtleties which I suspect will be 
of limited or no interest to most physicist readers.  Since I had to work them
out to be confident that the mathematics was rigorous, I thought I might as
well include them. 

\section{Notation and point of view}
	We attempt to stay as close as possible to traditional physics notation,
	reverting to notation more common in mathematics only when it seems 
	less ambiguous or complicated. 
	The inner product of vectors $v, w$ in a complex Hilbert space $H$  will be
	denoted $\lb v, w \rb$, with the physics convention that this be linear in
	the second variable $w$, and conjugate-linear in the first variable $v$.
	The norm of a vector $v$ will usually be denoted as 
	$ |v| := \lb v,v \rb ^{1/2}$, but 
	(with due notice to the reader) 
	this may occasionally be changed  to $|| v ||$ when dealing with Hilbert
	spaces of functions, to distinguish the  
	$L^2$ norm of a function from its absolute value. 
	 
	The mathematics of 
	quantum mechanics describes
	a (pure) state of a physical system as a ray in a complex Hilbert space, 
	a ``ray'' being defined as the set of all multiples
	$\alpha v$, $\alpha$ complex, of a nonzero vector $v$.  Thus a ray is 
	uniquely described by any nonzero vector in it, which may be normalized
	to norm 1 if desired.  In a context of detailed calculations, 
	normalization is often helpful, but otherwise it may merely introduce nuisance      
	numerical factors.  We shall normalize only where it seems helpful.  
	Also, we follow convention by speaking, regarding a vector $v$, of the  
	(pure) ``state'' $v$ instead of the more pedantic ``ray determined by $v$''.

	The projector to a subspace $E$ will typically be denoted $P_E$, in 
	place of the common but unnecessarily complicated 
	physics notation $\sum_i | e_i  \rb \lb e_i |$, where $ \{e_i\}$ is an 
	orthonormal basis for $E$.  
	When $E$ is the entire Hilbert space of states, $P_E$ is called the 
	{\em identity operator} and denoted $I := P_E$.   
	When $E$ is the one-dimensional subspace
	spanned by a vector $w$, we may write $P_w$ for $P_E$.  
	We shall make constant use of the formula for 
	$P_w$ when  $|w| = 1$: $P_w v = \lb w, v \rb w$.   
	Note, however, that under our convention, $ P_w = P_{w/|w|}$, so this 
	formula for $P_w$ only applies for $|w| = 1$.

	Our mathematical formulation of quantum mechanics follows that of 
	Chapter 2 of the  book of Nielsen and Chuang \cite{N/C}.  We shall consider
	only projective measurements (as opposed to the more general 
	``positive operator valued" measurements emphasized there).

	Given a system in a pure  state and a subspace $E$, 
the projective measurement associated with $E$
	answers the question:  ``After the measurement, is the state of the system
	in $E$ or the orthogonal complement $E^\perp$ of $E$?''.  (According to 
	usual interpretations of quantum mechanics, those are the only possibilities
	for such a projective measurement.)

	If before that projective measurement the system was in pure state $v$, 
	then after 
	the measurement it will be either in (unnormalized) state $P_Ev$ or $(I-P_E)v$, with
	respective probabilities $|P_E v|^2/|v|^2$ and $|(I-P_E)v|^2/|v|^2$
	(\cite{N/C} p. 87 ff.). 
	The measurement is sometimes said to 
	``project'' the pre-measurement state $v$ (which may be neither in 
	$E$ nor its orthogonal complement) into either $E$ or its orthogonal
	complement.%
	\footnote{This terminology is suggestive but slightly misleading because
	to ``project'' $v$ in this sense is not the same as applying a single
	projector to $v$.}
When $E$ is the one-dimensional subspace spanned by a 
vector $w$, the measurement is said to determine if the post-measurement
state is $w$, which for normalized states $v$ and $w$ will occur with 
probability $|P_w v|^2 = | \langle v, w \rangle |^2$. 
Note also that when $E$ is the entire pure state space, so that $P_E = I$,
the measurement has no effect on the system. 

More generally let
$E_1, E_2, \ldots $ be a finite or infinite collection 
of pairwise orthogonal subspaces with $\sum_i P_{E_i} = I$.  The collection
of projectors $\{P_{E_i} \}$ is called a 
{\em resolution of the identity}.   
With any such collection is associated a {\em projective measurement}
whose result is to ``project'' an initial (pre-measurement) pure state $v$ 
onto one of the $P_{E_i} v$ with probability $|P_{E_i} v|^2/|v|^2.$

The above refers to pure states.
We shall deal mainly with pure states, but it is sometimes difficult to discuss
measurements of pure states without introducing the slightly subtle 
concept of ``mixed'' state.  For simplicity of exposition, we shall review
this concept in a context less general than necessary, but as general
as we shall need it.

Let the unit vectors  $v_1, v_2, \ldots $ represent a 
(finite or countably infinite)
collection of pure states and 
$p_1, p_2, \ldots $ corresponding 
probabilities which sum to 1.  
Suppose we decide to choose randomly one of the pure states $v_i$ 
with probability $p_i$ and then perform some measurements on the chosen state.
Just before the random choice, we say that the system is in a {\em mixed} 
state which we'll denote by the traditional symbol $\rho$.  
Mathematically, a mixed state $\rho$ is described by a positive operator 
of trace 1, traditionally called a {\em density matrix}: 
for the above example, $\rho = \sum_i p_i P_{v_i}$.  
If after 
the random choice we measure to see if the (pure) state of the system  
lies in a subspace $E$,
the probability that it does lie in $E$ is $\tr (\rho P_E)$.
When all but one of the probabilities $p_i$ vanish, say the nonvanishing 
probability is $p_k =1 $, then 
$\rho = P_{v_k}$ and 
$\tr (\rho P_E) = \tr( P_{v_k} P_E ) =  |P_E v_k|^2/|v_k|^2,$ 
which is the same as the corresponding formula for a system in pure state 
$v_k$.%
\footnote{Since our formulation of ``mixed state'' does not exclude the
case of exactly one nonvanishing $p_i$, a pure state is technically a mixed
state.  Some authors use ``mixed state'' to denote what we
would have to call a ``non-pure mixed state''.}

For example, suppose the Hilbert space of pure states is two-dimensional with orthonormal
basis $e_1, e_2$, and consider a measurement to determine if a particle
with initial normalized state $v$ is (after the measurement) 
in state $e_1$ or $e_2$.  (This corresponds to a projective measurement
with respect to the resolution of the identity $P_{e_1}, P_{e_2}$.)
If we know that the measurement has been made but have not been told the result,
then from our point of view the system is in a mixed state:  it is in
pure state $e_1$ with probability $|P_{e_1} v|^2 = |\langle e_1, v \rangle|^2$ and in pure state
$e_2$ with probability $|\langle e_2, v \rangle |^2$.   But if we know the
result of the measurement, then the system is in a pure state, either $e_1$ or $e_2$,
according to the result.  A subtle aspect of the concept of ``mixed state''
is that the state of the system may depend on the observer:  an  observer
who knows the result of a measurement will ``see'' a pure state, but the state
will be mixed from the perspective of 
an observer who knows that a measurement has been made (or will be made) 
but does not know the result.

\section{The problem which ``weak'' measurements address}
Consider the following trial of an experiment which is to be repeated
many times.
A quantum system $S$ is prepared in a given initial state $s_{in}$. 
We have an observable (Hermitian operator) 
 $A: S \rightarrow S$.  For simplicity, assume that $A$ has a complete
set of eigenvectors.  We would like to measure the expectation 
of $A$ in the state $s_{in}$, namely $ \lb s_{in}, A s_{in} \rb$, 
{\em without changing the state of $S$} (from $s_{in}$ to something else).

What was just proposed may seem impossible because
measuring $A$ constitutes a projective measurement 
which will project $s_{in} $ onto one of the
eigenvectors of $A$ (thus changing it, unless $s_{in}$ 
happened to be an eigenvector
to start).  However, we did not say we needed to measure $A$, but only
its expectation in the state $s_{in}$, which is not quite the same thing.
And we are willing to weaken the requirement of not changing $s_{in}$ to
only changing $s_{in}$ negligibly.

Why would we want to do this?  Well, we might want to perform a subsequent
experiment on $s_{in}$ and speak of ``the expectation of $A$ in the 
state $s_{in}$ given that'' the subsequent experiment yielded a particular
result.  Classically, this would be called a conditional expectation 
of $A$ given the particular result of the subsequent experiment. 

In practice, the ``subsequent experiment'' is usually a so-called
``postselection'' to a given final state $s_f$.  
We measure $\lb s_{in}, As_{in} \rb$ while changing $s_{in}$ neglibibly 
and then perform a projective measurement to see if the final state 
is $s_f$ (as described in a preceding section).  If the final state {\em is}
$s_f$, we record the measurement of 
$\lb s_{in}, A s_{in} \rb$; 
otherwise we discard it.  The average of all the recorded measurements  
$\lb {s}_{in}, A {s}_{in} \rb$ is taken as (a good approximation
to) the average value of $A$ in 
all states $s_{in}$ which  project to $s_f$ in the final step.  Classically,
this would be called the conditional expectation of $A$ given that  the
final measurement yielded $s_f$.

The reader who feels a sense of unease at this proposal should be congratulated
on his%
\footnote{or her, of course.  We follow the long-standing 
and sensible grammatical convention that
either ``his'' or ``her'' means ``his or her'' or ``her or his'' 
in contexts like this.
We find it distracting to alternate ``his'' with ``her'', and make no 
attempt to equalize the number of each.}  
perspicuity.  Indeed, serious objections can be raised to the proposal,
but we are presenting it as it seems typically 
(though usually not so explicitly)
presented in the research literature. 
Objections will be discussed later.
Right now we are trying to present a rapid overview.

Part of the above proposal {\em is} both sound and experimentally feasible.  
A seminal paper
of Aharonov, Albert and Vaidman \cite{AAV} cleverly 
expands a theory of measurement
of von Neumann to accomplish measurement of $\lb s_{in}, A s_{in} \rb$
while negligibly changing the state of the system. 

They couple the original system $S$ to an auxiliary ``meter''
system $M$ in such a way that a particular ``meter'' observable in the
meter system has the same mathematical expectation as $A$,
and in addition the coupling between the meter and the original
system $ S$ is so weak that measuring the meter observable 
negligibly affects the state of $S$.   

This involves a tradeoff between weakness of the coupling and accuracy
of individual meter measurements.  A weaker coupling generally entails
wider dispersion of the meter measurements.  The average of a large number of 
meter measurements will approximate $\lb s_{in}, A s_{in} \rb$, but a very
large number may be required when the coupling is very weak. 

\section{A simple finite-dimensional model} 
\subsection{Introduction of the model}
We present a simple model which illustrates how weak measurements 
can be performed.  Let  $S$ denote the Hilbert space of the system 
which is the primary object of study, and  $M$ the Hilbert space of 
an auxiliary ``meter'' system.  We will take both $S$   and $M$ to be 
two-dimensional
 with orthonormal bases $s_0, s_1$ for $S$  and $m_0, m_1$ for $M$. 
No confusion should result if we use the same symbol $S$ for the physical system
itself and for its Hilbert space, and similarly for $M$.

The Hilbert space of the composite system comprising both $S$ and $M$ 
is the tensor product $S \tnr M$, which has the orthonormal
basis $\{s_i \tnr m_j \,|\  0 \leq i,j \leq 1 \}.$%
\footnote{This is a note for the inexperienced reader, and also
to bridge the gap between the traditional physics notation of 
our primary reference \cite{N/C} and our slightly different notation
(which we think simpler for the kind of calculations which
we shall be doing).  
\s
One definition of ``tensor product'' of the above $S$ and $M$ (the simplest but not
necessarily best definition) is as a $2 \times 2 = 4$ dimensional
Hilbert space with orthonormal basis denoted as above and with 
a binary operation $\tnr$ which assigns to any $s \in S$ and $m \in M$ an
element denoted $s \tnr m$ of $S \tnr M$,  $(s,m) \mapsto s \tnr m$, 
such that $\tnr$ satisfies the distributive rules of algebra:  
for any scalars $\sigma_i, \mu_j$,
$$ (\sigma_0 s_0 + \sigma_1 s_1) \tnr (\mu_0 m_0 + \mu_1 m_1) = 
\sigma_0 \mu_0\, s_0 \tnr m_0 + \sigma_0 \mu_1 \, s_0 \tnr m_1 
+ \sigma_1 \mu_0\, s_1 \tnr m_0 + \sigma_1 \mu_1 \, s_1 \tnr m_1 .
$$ 

Typical physics notation for $s_i \tnr m_j$ might be something like
$|i\rb_S |j\rb_M$.  In this notation, the tensor product symbol is
omitted.  It could be omitted in our notation, too, but it seems to 
enhance readability.  
}
If the system $S$ is in state $s$  and system $ M$ in state $m$,
then the composite system is in state $s \tnr m$, and 
conversely.  
Not every state of $S \tnr M$
can be written in the form $s \tnr m$.  Those which can be written in this way are are called {\em product states}, 
and those which cannot are called {\em entangled}.  
An example of a state which can be proved to be entangled 
is $s_0 \tnr m_0 + s_1 \tnr m_1.$
\subsection{The states of $S$ and of $M$ derived from the
state of  $S \tnr M$}

When the composite system $S \tnr M$ is in an entangled pure state, 
each individual system $S$ or $M$ is in mixed state, which is never pure.  
This follows from  the general rule (\cite{N/C}, p. 107) for 
passing from a mixed state $\rho$ (which could be pure)
of the composite system $S \tnr M$ 
to states of its factors $S$ and $M$. 
The rule is that the state of $S$ is
obtained by taking the partial trace with respect to $M$ of $\rho$, denoted
$\trM \rho$. 
A consequence of this rule is that  
when $S \tnr M$ is in a pure state, 
a (projective) measurement in $ M$ can affect the state of $ S$ 
if and only if 
the state of $S \tnr M$ is entangled.%
\footnote{It is immediate that if a pure state of $S \tnr M$ is {\em not}
entangled (i.e., is a product state $s \tnr m$), then a measurement in $M$ cannot
affect the state $s$ of $S$.  To see that if a pure state {\em is}
entangled, then a measurement in $M$ {\em will} change the state of $S$,
let $m_0, m_1, \ldots $ be an orthonormal basis for $M$.
Then a general normalized pure state $r$ of $S \tnr M$ can be written 
uniquely as
$r =  \sum_i s_i \tnr m_i$, with $s_i \in S$ satisfying $\sum_i |s_i|^2 = 1$.  

A projective measurement in $M$ with respect
to the basis $\{m_i\}$ corresponds to a projective measurement in $S \tnr M$
with respect to the resolution of the identity $\{I \tnr P_{m_i} \}$.  
This results in outcome $m_k$ occuring with probability $|s_k|^2$, and
when $m_k$ does occur, the resulting state is the pure state 
$(I \tnr P_{m_k}) r = s_k \tnr m_k$.  That is, the measurement changes $r$ 
into $s_k \tnr m_k$ with probability $|s_k|^2$.  In order for this not 
to be an actual change, $r$ must be a multiple of  
$ s_k \tnr m_k$, which says that
$r$ is a product state. 
}

Appendix 1 reminds the reader of the definition and properties 
of the partial trace.  One very useful calculation done there encapsulates
most of the properties of the partial trace which will be used in our 
discussion of weak values.  It goes as follows.  Let $\{m_i\}$ be an 
orthonormal basis (finite or infinite) for $M$ and $r$ a pure state in
$S \tnr M$ with $|r| = 1$  
It is easy to show that $r$ may be uniquely written as 
\beq
\lbl{eqS10}  
r = \sum_i s_i \tnr m_i \q \mbox{with $s_i \in S$ and $\sum_i |s_i|^2 = 1$}. 
\eeq
(The $s_i$ need not be orthogonal.)
Considered as a mixed state, $r$ is represented by the density matrix  
$P_r$, and the corresponding mixed state of $S$ is $\trM P_r$.
The calculation yields
\beq
\lbl{eqS20}
\trM P_r =  \sum_i |s_i|^2 P_{s_i} \q, 
\eeq  
which explicitly exhibits the mixed state of $S$ corresponding to the pure 
state $r$ of $S\tnr M$ as a convex linear combination of pure states
of $S$.
This rule for passing from a pure state of $S \tnr M$ to a mixed state
of $S$ will be illustrated by an example which will also  
illustrate other subtleties.

Take both $S$  and $M$ to be two-dimensional 
with orthonormal bases $s_0, s_1$ and 
$m_0, m_1$, respectively. 
Suppose we perform the following experiment a large number
of times.  Start with the composite system $S \tnr M$  
in the normalized entangled state 
$$e:= (s_0 \tnr m_0 + s_1 \tnr m_1)/\sqrt{2}.$$  
Then perform a projective
measurement in the meter system to determine if it is in state $m_0$ or $m_1$. 
(This corresponds to a projective measurement in the composite system with
respect to the resolution of the identity 
$I \tnr P_{m_0},\  I \tnr P_{m_1}$.) 
The result of the projective measurement will be  $m_0$ with probability 
$|(I\tnr P_{m_0})e|^2 = 1/2$, and in this case 
the post-measurement state will be the product state 
$(I\tnr P_{m_0})e = s_0 \tnr m_0/\sqrt{2}$.  (The observable $I \tnr P_{m_0}$ 
corresponds to measuring $P_{m_0}$ in the meter system and nothing in
the $S$ system.)  The corresponding state of $S$ expressed as a density matrix
is $\trM P_{s_0 \tnr m_0} = P_{s_0}$, which expressed as a vector
state is $s_0$. 
Similarly, it will be found to be $m_1$ with probability 1/2, in which case
the post-measurement state will be $s_1 \tnr m_1$, with $S$ in state $s_1$. 

This seems to imply that a measurement in the meter system can affect the state
of $S$, and indeed it can, according to what seems the generally accepted
physical interpretation of the mathematics.  (This seems 
the interpretation of \cite{N/C} and is also our interpretation.) 
But the way in which it affects the state of $S$ is subtle and worth
further examination.  

In principle, the two systems $S$ and $M$ could be spacelike separated, so
that, according the principle of relativistic locality, a measurement in $M$ at 
a particular time (in some Lorentz frame) could 
{\em not} affect a measurement in $S$  at the same time.  
That implies that if the $S$ system can be said to have
a definite state, that state {\em cannot}
be affected by a particular $M$ measurement.  
This seems to contradict the interpretation which we adopted in the preceding
paragraph.  

The contradiction is one manifestation of a pervasive 
tension between ``realism''
and ``locality'', two philosophical concepts which are typically only 
vaguely defined in the literature.%
\footnote{Including {\em this} literature!  We are not trying to prove a theorem
here, but to communicate a point of view.  An attempt at careful definitions 
(which would not be easy) would only be a distraction.}
It would seem that if we accept that
system $S$ must have a definite state (one aspect of ``realism'') and that a measurement
in a spacelike separated $M$ at a particular time can affect that state at that
time, then the principle of locality cannot hold.

The most common resolution of this contradiction seems to be to abandon
``locality'' as a general principle (i.e., reject the proposal 
that measurements in the meter system 
at a particular time cannot influence the $S$ system at the same time)  
and replace it by the weaker statement that measurements in the 
meter system cannot be used to send superluminal messages (i.e., messages which
travel faster than light) to the $S$ system.

That removes the contradiction, but leaves one a little uneasy that 
although there seems no obvious way that the mathematics of quantum mechanics
could facilitate superluminal communication, nevertheless some clever person
might someday find a way to do it. 

We will accept this resolution because without something like it, we could
not use the accepted physical interpretations of the mathematics 
of quantum mechanics.  The resolution applies as follows to the experiment 
introduced above
of measuring for $m_0$ or $m_1$ in  the meter system.

Perform the experiment
 a large number of times, and divide the outcomes 
into two classes, those which resulted in $m_0$ (Class 0) and those resulting in
$m_1$ (Class 1).  Then any measurements made in $S$ for the Class 0 states 
will have identical statistics to measurements made in a single system $S$
(i.e., forgetting entirely about $M$) in state $s_0$, and similarly the Class 1 
states will have statistics identical to those of system $S$ in state $s_1$.  
The combined classes will have statistics identical to those of a mixed 
state which is in state $s_0$ with probability 1/2 and $s_1$ with the same
probability.

Although the $S$ observer may not realize it, half the time he is measuring
in state $s_0$, and half the time in $s_1$.  Considered as a density
matrix (trace one positive operator) on $S$, his state is 
$(1/2) P_{s_0} + (1/2) P_{s_1} = I/2$ {\em until he is informed of the result
of the meter measurement}. 
If the measurement in the meter system results, say, in outcome $m_0$,
and if the $S$ observer is informed of this fact, then he will condition
his future calculations on this fact.  When so conditioned, his state 
becomes $P_{s_0}$ (considered as a density matrix) or $s_0$ (considered
as a vector).  
 
Only in this weak statistical sense, 
can the existence or nonexistence of a measurement
in $M$ (and its result) influence the state of $S$.  But for $S$ to
perceive this influence seems to require ``classical'' (i.e., not faster
than light) communication between $M$ and $S$.  

\subsection{Introduction to ``weak'' measurements} 
We have noted that when $S \tnr M$ is in a pure state $s \tnr m$,
a projective measurement in $M$ cannot affect the state of $S$.
More generally, the effect of a measurement in $ M$ 
on the state of $ S$ is expected to be arbitrarily small if the 
pre-measurement
pure state of $S \tnr M$ is close enough  to a product state.  To make
this precise, we would have to commit ourselves to definitions of
``effect of a measurement in $ M$'' and ``close enough to a product
state''.  When the Hilbert spaces $S$ and $M$ are finite dimensional, 
so are the linear spaces in which the mixed states reside, and since
all norms on a finite dimensional normed linear space are equivalent, 
closeness can be measured in any convenient norm.

For example, we could measure the distance between two mixed states $\rho, \sigma$ using the trace norm:
$|| \rho - \sigma ||_{tr} := \sqrt{\tr(\rho-\sigma)^2}.$ 
The reader who wants 
a definite definition may use the one just given, 
though other norms may be easier to calculate in specific contexts.

For pure states
in $S \tnr M$, the reader will probably find it easier 
to think in terms of the Hilbert space norm instead of
the trace norm.  (We didn't formulate our definition in terms of the 
Hilbert space norm because we needed it to apply also to mixed states,
in order to talk about the ``effect of a measurement in $M$ on the 
(generally mixed) state of $S$''.) 

In infinite dimensions, more care would be necessary.  
These issues seem not to have been considered in the literature
on weak measurements.
We do not pursue them here
because our aim is to explain the concept of weak measurements in the 
context in which they historically arose, in which technical mathematics
would only be a distraction.  

{\em For the rest of this section, we
assume that all Hilbert spaces occurring are finite dimensional.}
This implies that all Hermitian operators occurring can be written as
a linear combination (with coefficients the eigenvalues of the operator)
of orthogonal projectors.

With these preliminary observations out of the way, we can describe
the idea of weak measurement in a more precise way.  Suppose we are given 
a reproducible, normalized pure state $ s $ of $S$. 
By ``reproducible'' we mean that we have an apparatus that can produce
any number of replicas of this state.%
\footnote{
This does not contradict the supposed impossibility of cloning 
quantum states (under certain restricted hypotheses a ``no-cloning''
theorem can be proved) because we do not require to be able to copy 
{\em any} unknown state given to us, 
only that we have a device which can reliably produce
some particular state in which we are interested. 
} 
Suppose we are also given
an observable (Hermitian operator) $A$ on $S$.  Our goal is to determine
the expectation $\lb s, As \rb$ of $A$ to arbitrary accuracy
by making measurements only in $ M$ which have arbitrarily small
effect on the pre-measurement state $s$ of $S$.  
We shall call these ``weak'' measurements.%
\footnote{In the literature, the term
``weak measurement'' may sometimes refer to experiments which involve
not only weak measurements in our sense, but also ``postselection'' 
to be described below.}

We now describe a simple way to do this.  Let $m_0, m_1$ be an
orthonormal basis for $M$.  Let $\epsilon > 0$ be a real parameter 
which will measure the strength of the effect of a ``meter'' measurement in
the $m_0, m_1$ basis on the state of $S$.  
Here we are using the descriptive phrase ``meter measurement in the 
$m_0, m_1$ basis'' 
as shorthand for a projective measurement relative to
the resolution of the identity $\{I \tnr P_{m_0}, I \tnr P_{m_1}\}$ in
$S \tnr M$. 

First we consider the special case in which $A$ is {\em both} Hermitian
and unitary (as are the Pauli matrices, for example).  
Let $s$ be a normalized pure state of $S$, and $m_0, m_1$ an orthonormal
basis for $M$.  
Start with the pure state of $S \tnr M$
\beq
\lbl{eq80}
r := r(\epsilon) := s \tnr m_0 \sqrt{1-\epsilon^2} + As \tnr m_1 \epsilon
\q.
\eeq
The assumptions that $A$ is unitary and $s$ is normalized imply 
that $r$ is normalized:  $|r| = 1$. 

From the point of view of $S$, this state
is very close to the given state $s$ when $\epsilon$ is small.  
More precisely, the partial trace with respect to $M$ of $P_r$, $\trM P_r$,
is arbitrarily close to $P_s$ for $\epsilon$ sufficiently close to 0. 

Let 
\beq
\lbl{eq90}
 B := 
\left[
\begin{array}{ll}
B_{00} & B_{01} \\
B_{10} & B_{11}
\end{array}
\right]
\eeq
be the matrix with respect to the $m_0, m_1$ basis of a Hermitian
operator $B$ on $M$. Our main conclusion will require the assumption
that $B_{00} = 0$, but we initially include an arbitrary $B_{00}$ 
for purposes of later discussion.  

A short calculation reveals that
$$
\lb r(\epsilon), (I \tnr B) r(\epsilon) \rb = B_{00} (1 - \epsilon^2) 
+ 2 \epsilon \sqrt{1-\epsilon^2} \lb s, As \rb \Re (B_{10})    
+ \epsilon^2 \lb As, As \rb B_{11} \q. 
$$
Under the assumption $B_{00} = 0$, 
\beq
\lbl{eq100} 
\lim_{\epsilon \goesto 0} \frac{\lb r(\epsilon), (I \tnr B)r(\epsilon) \rb}
{\epsilon} \ =\  2 \lb s, As \rb \Re (B_{10}) 
\q. 
\eeq
This says that when $\Re B_{10}  \neq 0$, 
measuring the average
value of $B$ in $M$  and normalizing appropriately (by dividing by $\epsilon$)
will approximate the average value of $A$ in state $s$ of $S$ 
up to an inessential constant factor. 
Moreover, we shall show in a moment that this approximation is obtained with
negligible change (for small $\epsilon$) in the state $s$ of $S$.

Although we assumed that $A$ was unitary in order to make the algebra
simple, similar weak measurement protocols
can easily be obtained for arbitrary Hermitian operators $A$ (on a finite
dimensional $S$).
The following are two possible approaches.
\begin{enumerate}
\item
If $P$ is a projector, then $U := 2P-I$ is unitary (and Hermitian), 
and $ P = (U+1)/2$.
Since we showed above how to
obtain $\lb s, Us \rb$ to arbitrary accuracy by measurements in $M$, 
the simple algebraic transformation $\lb s, P s \rb
= \lb s, Us \rb /2 + 1/2 $ gives us $\lb s , Ps \rb$ with the same 
(negligible for small $\epsilon$) 
disturbance of the state $S$.  
The case of general Hermitian $A$ is then obtained by writing $A$ as a linear 
combination of projectors.  
\s
This approach has the advantage of conceptual simplicity, 
but may be inconvenient in specific
applications. 
\item
The case of general Hermitian $A$ can also be obtained  by normalizing
\re{eq80}, i.e., replacing $r(\epsilon)$ in \re{eq80} by
the normalized state
\beq
\lbl{eq105}
\tilde{r}(\epsilon) := \frac{r(\epsilon)}{|r(\epsilon)|}
=  \frac{s \tnr m_0 \sqrt{1-\epsilon^2} + As \tnr m_1 \epsilon } 
{\sqrt{1 - \epsilon^2 + \epsilon^2 |As|^2}}
\q.
\eeq
Though the algebra is slightly messier, 
\re{eq100} with $\tilde{r}(\epsilon)$ in place of $r(\epsilon)$ 
is still obtained:  
\beq
\lbl{eq107}
\lim_{\epsilon \goesto 0} 
\frac{\lb \tilde{r}(\epsilon), (I \tnr B)\tilde{r}(\epsilon) \rb}
{\epsilon} \ =\  2 \lb s, As \rb \Re (B_{10}) 
\q. 
\eeq 
\end{enumerate}

Taking $B_{10} := 1/2 =: B_{01}$,
we see that measuring the average value of $B$  relative to $\epsilon$ 
will yield the average value of $A$ in state $s$ to arbitrary
accuracy.  If the measurement of $B$ 
has arbitrarily small effect on the
state of $S$ for sufficiently small $\epsilon$, then 
$B$ ``weakly'' measures $\lb s, As \rb$. (This is essentially our definition
of ``weakly measures'', which will be formalized in Definition 1 below.)

It is not hard to see that for small $\epsilon$, measuring $B$ 
 does not appreciably affect the state of $S$. 
For small $\epsilon$, $\tilde{r}(\epsilon)$ is arbitrarily
close to $s\tnr m$, and the corresponding mixed state of $S$, namely
$\trM P_{\tilde{r}}$, is arbitrarily close to $\trM P_{s\tnr m} = P_s$ 
(which is the density matrix equivalent of vector state $s$).  
Here we have used the facts that in finite dimensions, $r \mapsto P_r$
($|r| = 1$)
and  
$\rho \mapsto \trM \rho$ are continuous relative to the topologies induced 
by any norms.

Let $b_0, b_1$ be orthonormal eigenvectors of $B$.
The measurement of $B$ will project $\tilde{r} (\epsilon)$ onto 
either $(I \tnr P_{b_0})\tilde{r}(\epsilon)$ or 
$(I \tnr P_{b_1}) \tilde{r}(\epsilon)$.  For small $\epsilon$, this
projection will be arbitrarily close to either 
$(I \tnr P_{b_0})(s \tnr m) = s \tnr P_{b_0} m$ or
 $ s \tnr P_{b_1}m$, respectively.  
Taking the first alternative for convenience, we may conclude that 
the post-measurement state of $S$, namely the density matrix 
$\trM P_{(I \tnr P_{b_0}) \tilde{r}}$, is arbitrarily close to
$\trM P_{(I \tnr P_{b_0})(s \tnr m)} = \trM P_{s \tnr P_{b_0} m}
= P_s$.  This shows that for small $\epsilon$, measuring $B$ has arbitrarily
small effect on the state of $S$.

Measuring the average value of $A$ without significantly changing the 
state of $S$ may
 may seem too good to be true, but close scrutiny reveals that a tradeoff
is involved.  Measuring the average value of $B$
requires measuring a very small quantity (on the order of $\epsilon$). To
obtain a reliable average value for $B$ we will need to average a large number
of individual observations.  The smaller $\epsilon$, the larger the number
of observations required.  (From general statistical theory, we expect 
that the required number of observations should scale like $1/\epsilon^2$.) 

In an ideal situation in which measuring $B$
produces only random errors, the random errors will average to zero, and 
the procedure described will be feasible.  But systematic errors need not
average out and can make the weak measurement infeasible when $\epsilon$
is too small.  This drawback, or tradeoff, is implicit in all ``weak
measurement'' schemes which we have seen. 

For example,
in a Stern-Gerlach experiment to measure the spin of particle, the slightest
misalignment of the magnets can produce a systematic error which will cause
the limit $\epsilon \goesto 0$ in \re{eq100} to fail to exist.  Mathematically,
this shows up in the necessity of the assumption $B_{00} = 0$.  In a physical
experiment, there is generally no way to assure that $B_{00}$ will be
 {\em exactly} zero.  But if $B_{00}$ is small, we can hope to obtain meaningful
experimental results consistent with \re{eq100} 
when $\epsilon$ is not too small.  

Since we shall often refer to ``weak measurements'', we offer the following
physical definition of such measurements.  
\begin{definition}
\lbl{def1}
Let $S$ be a quantum system and  $A$ an observable on $S$.
Let $s$ be a state of $S$ of which we can make an arbitrary number 
of copies. 
A {\em weak measurement protocol} is a procedure which can determine the
expectation $\lb s, As \rb$ of $A$ in the state $s$ to arbitrary accuracy
while perturbing each copy of $s$ used in the procedure 
by an arbitrarily small amount.  
\end{definition}

Weak measurements are of particular interest
when we want to use the (copies of the) state $s$ used in the weak measurement
for future experiments.  If we had measured
observable $A$ in state $s$ in the normal way, it would project $s$ onto one of the eigenvectors
of $A$, thus changing $s$ to that eigenvector. 

A kind of ``future experiment''  which has received great attention
in the literature is ``postselection''.  Indeed, interest in weak
measurements stems from the seminal paper of Aharonov, Albert, and Vaidman 
\cite{AAV}
which considers weak measurements in the context of postselection and
coins the term ``weak value of a quantum variable''. 

Let $f  $ be a desired ``final state'' in $S$.%
\footnote
{
Previously, the initial state was called $s_{in}$ and the final state $s_f$
to correspond more closely to the notation $\psi_{in}$  and $\psi_f$
commonly used in the literature.  But both notations are unnecessarily 
complicated.  The only additional information contained in $s_f$ as opposed 
to $f$ is an implicit suggestion that it probably refers to a state 
in $S$.  But since the final state $f$ always occurs in expressions like
$f \tnr m \in S \tnr M$ which explicitly imply the suggestion, this additional
information is redundant and complicates the typesetting. 
\s
While on the subject, we also warn 
the reader that earlier versions of this paper used $q$ instead of $f$ or 
$s_f$ for the final state.  Should the reader encounter a $q$ in the context
of a final state,
it is probably a typo. 
} 
A ``postselection'' experiment 
follows a weak measurement protocol by a projective measurement relative
to the resolution of the identity 
$P_f \tnr I_M , (I_{S}-P_f) \tnr I_M) $, where the subscripts on the
various identity operators denote the spaces on which they act.  (In the
future, we shall generally omit the subscripts, writing, for example,
$P_f \tnr I,  (I-P_f) \tnr I$ for the above.)  

Suppose the state of $S \tnr M$ 
after the weak measurement protocol is a pure state
(an atypical situation, but the easiest to discuss).
Then informally, the postselection asks (and answers) the question: 
``Is the state of 
$S \tnr M$ after postselection in $[f] \tnr M$ or $[f]^\perp \tnr M$, where
$[f]$ denotes the subspace spanned by $f$, and $[f]^\perp$ its orthogonal
complement.  If the postselected state of $S\tnr M$ is in $[f] \tnr M$,
then it is  of the form $f \tnr r$ with $r \in M$, and the corresponding 
state of $S$ is $f$.  Otherwise, the state of $S $ is a mixture of pure states
which are orthogonal to $f$.  

``Postselection'' discards cases in which
the postselected state of $S$ is not the pure state $f$.  
This amounts to replacing 
probabilities by probabilities conditional on successful (i.e., the
postselected 
 state of $S$ is $f$) postselection.

In the more typical case in which the state of $S \tnr M$
after the weak measurement protocol is a mixed state $\rho$ (a density matrix), 
then after the postselection it will be either in state 
$(P_f \tnr I) \rho (P_f \tnr I)  / 
\tr((P_f \tnr I) \rho (P_f \tnr I  )) $ or
$((I - P_f) \tnr I) \rho ((I - P_f)  \tnr I)  / 
\tr(((I - P_f) \tnr I) \rho ((I - P_f) \tnr I))),$ 
so the postselection may be regarded as resolving the question:  
Does the postselected state
$\sigma$ satisfy 
$((I - P_f) \tnr I) \sigma = 0$ (successful postselection)   
or $(P_f \tnr I) \sigma = 0$?

The cumbersome language of mixed states
can usually be finessed in the context of weak measurements 
in finite dimensions
by noting that a sufficiently accurate 
weak measurement protocol
starting with initial state $s \in S$
leaves $S$ in a state close to $s$.  It is routine to show that this implies 
that the state of $S \tnr M$ is close to (the density matrix analog 
$P_{s \tnr r}$ of) 
a pure product state $s \tnr r$.  Then one can apply the simpler
discussion previously given to the pure state $s \tnr r$. 

In the literature, the result of measuring the average value of
an observable in a state $s$ by a procedure which negligibly disturbs 
the state (such as \re{eq100} above) followed by conditioning on 
successful postselection to a final state $f$  
is said to result in a ``weak value''
of the observable.  If in the above example we choose 
$B_{10 } := 1/2 =: B_{01} $, we obtain the so-called ``weak value''%
%
%
\begin{eqnarray}
\lbl{eq110}
\lim_{\epsilon \goesto 0} 
\frac{ 
\lb \tilde{r}(\epsilon), (P_f \tnr B)\tilde{r}(\epsilon) \rb 
}
{\epsilon \lb \tilde{r}(\epsilon), (P_f \tnr I) \tilde{r}(\epsilon) \rb }
&=&  
\frac{
\lb As, P_f s \rb /2   + \lb s, P_f As \rb) /2 
}
{\lb s, P_f s\rb } \nonumber\\
&=& 
\frac{\lb f,s \rb \lb As, f \rb /2  + \lb s, f \rb \lb f, As \rb /2
} 
{ \lb f, s \rb \lb s, f \rb} \nonumber\\
&=& 
\Re \frac{ \lb f, As  \rb}
{\lb f, s \rb} 
\q. 
\end{eqnarray}
The mathematical procedure may be understood as follows. 
Asking the postselection question ``Is $S$ in state $f$ or in a mixture of
states orthogonal to $f$?'' 
is the same as measuring $P_f \tnr I$, and the expectation of $P_f \tnr I$ in
state $\tilde{r} (\epsilon)$ is the proportion of ``yes'' answers (in the limit
of a large number of measurements).  Measuring $P_f \tnr B = (P_f \tnr I)
(I \tnr B)$ corrsponds to simultaneously measuring $B$ and postselecting to 
$f$; its average is the average of the observable whose value is the value
of $B$ for successful postselections and zero otherwise.  The conditional
expectation of $B$ given successful postselection is the average of the just 
mentioned observable $P_f \tnr B$ divided by the probability of successful
postselection.  This conditional expectation is $O(\epsilon)$, so we normalize
by dividing by $\epsilon$.%
\footnote{In the narrative, we have been speaking of measuring $I \tnr B$ then
postselecting to $f$.  We are modeling this as measuring $I \tnr B$ and 
simultaneously postselecting (which is possible because $P_f \tnr I$ and 
$I \tnr B$ commute), which is the same as measuring $P_f \tnr B$.  The 
result of this simultaneous measurement is the result of the $I \tnr B$
measurement if the postselection was successful, and zero otherwise.
\s
The narrative could be revised to speak of simultaneous measurement of $B$
and postselection, but it seems easier to think about first measuring $B$, then
postselecting immediately after.
}

The  term ``weak value'' is not used consistently in the literature. 
The seminal paper \cite{AAV} of Aharonov, Albert, and Vaidman introduced the concept of  
``weak value of a quantum variable'' in a way conceptually similar 
to the above, but using a very different ``meter system'' along with a 
very different measurement procedure (which will be discussed in
the next section).  They identified 
$\lb f, As \rb /\lb f, s \rb$ (a quantity which in general need not
be real) as their ``weak value''.  Other authors (e.g., \cite{LS})
obtain the real part of this as their ``weak value'' 
(as we did above, using different methods).  
Some authors seem to {\em define} 
the ``weak value'' of an observable $A$ to be one of these expressions.
Most of the literature gives the
impression that any ``weak measurement'' (in the sense
of Definition 1) followed by postselection will result in a ``weak value'' 
$\Re (\lb f, As \rb / \lb f, s \rb)$, no matter what the measurement procedures.
One of the purposes of this note is to dispel this belief; a counterexample
will be given below.

At this point we have to warn the reader that in order to continue with the
description of the weak measurement process as typically presented in the 
literature (e.g. as in \cite{AAV}), we need to temporarily use language 
which we think questionable, as will be discussed in detail later.  
Suppose we start
with a state $s \in S$, construct $\tilde{r}(\epsilon) \in S \tnr M$ 
for suitably small $\epsilon$, measure $B$ in $M$ (i.e., measure 
$I \tnr B$ in $S \tnr M$), and then postselect to $f \in S$. 
We do this as many times as necessary to obtain a reliable average
of the results (i.e., measurements of $B$ followed by successful postselection).  
That average is taken as an estimate of  
the expectation of $B$ conditioned on postselection to $f$, which (after
normalization by division by $\epsilon$)
in turn is used as an estimate of {\bf the expectation of $A$ in the state
$s$,  conditioned on the postselection of $f$}. 
The emphasized phrase is problematic, but to the best of my understanding,
it accurately reflects the meaning of less precise language 
typical in the literature.

Why is the emphasized phrase problematic?  After all, conditional expectations
are uncontroversial in classical probability theory.  Insight can be obtained
by digressing to review classical conditional expectations on finite
probability spaces.

Let $\Sigma$ be a finite probability space, $\Delta$ a subset of $\Sigma$,
and $X$ a (real-valued) random variable on $\Sigma$ taking on values
$x_1, x_2, \ldots, x_m$.  Then the expectation $\Exp(X)$ of $X$, is defined by
$$
\Exp(X) := \sum_{i=1}^m x_i\, p(X = x_i) \q,
$$
where $p$ denotes probability.  Note that  $\Exp(X)$ is necessarily 
a convex linear combination of the values $x_i$ of $X$.
Conditioning on $\Delta$ means passing to a new probability space whose
underlying set is $\Delta$, with new probabilities $p_\Delta (\cdot)$
obtained by dividing 
original probabilities by $p(\Delta)$: for any subset $K$ of $\Delta$,
the new probability $p_\Delta(K) := p(K)/p(\Delta)$.
Relative to this new probability
space, the new expectation of $X$, denoted $\Exp(\,X\, |\, \Delta ,)$, is
$$
\Exp(\ X\,|\,\Delta\,) := \sum_{i=1}^m x_i\, p_\Delta(X = x_i) \q,
$$
which is again a convex linear combination of the values of $X$.  

The point is that in ordinary probability theory, 
{\em conditional expectations of an observable ({\rm i.e., random variable)} 
are always convex linear combinations of the values of the observable}.  
In quantum mechanics, the possible measured values 
of an observable on a finite-dimensional Hilbert space are its
eigenvalues.  But \re{eq110} need not be a convex linear combination of
the eigenvalues of $A$.  So it seems questionable to think of \re{eq110}
as a conditional expectation (as seems the usual interpretation in
the literature).   

This is emphasized by the provocative title of Ahaharonov, Albert, 
and Vaidman's seminal paper \cite{AAV}:  
``How the Result of a Measurement of a Component of the Spin of a Spin-1/2 Particle
Can Turn Out to be 100''.   The measurement they seem to be talking about is 
not a single measurement (which certainly could be far from expected 
if  experimental error is large), 
but a measurement of the {\em average} value of the spin, 
conditioned by a postselection.  The authors seem to believe that they have
explained how the average postselected value of a spin-1/2 
measurement can be 100.  
We are not convinced that the measurements 
they describe are measurements of the spin, conditioned on successful
postselection, as will be elaborated below.  

Although the widely accepted \re{eq110} surely 
describes {\em some} quantum measurement, we see no reason that it
should correspond to 
conditioning a measurement of the average
value of $A$ on postselection to $f$.  What it {\em does} 
correspond to for our 
toy model is given precisely by the left side of equation \re{eq110}, namely
conditioning the {\em meter measurement} on postselection to $f$.  The sentence
\begin{quote} 
1.  ``The average value of the normalized (i.e., by division by $\epsilon$) 
meter measurement equals 
the average value of $A$'' 
\end{quote}
is true (in the limit $\epsilon \goesto 0$) , but the sentence
\begin{quote}
2.  ``The average value of the normalized meter measurement conditioned 
on postselection to $f$ equals 
the average value of $A$ conditioned on postselection to $f$''
\end{quote}
is either false, meaningless, or tautological, depending on 
how it is interpreted.

In order to speak meaningfully of 
``the average value of $A$ conditioned on postselection
to $f$'', we need to say how this quantity is measured.  If we measure
$A$ in $S$, successfully postselect to $f$, and average the results, 
we do {\em not} necessarily
obtain \re{eq110}, $\Re (\lb f, As \rb / \lb f, s \rb)$. 

Physically, this is because measuring $A$ can significantly disturb the 
original state $s$ of the system. 
To explicitly calculate what happens,  
suppose that $A$ has two eigenvalues $\alpha_1, \alpha_2$ with
corresponding normalized eigenvectors $a_1, a_2$, and take $|s| = 1$.  
After $A$ is measured, $S$ is in state $a_i$ with probability 
$|\lb a_i, s \rb |^2$, $ i = 1,2$. Subsequently, the postselection succeeds with
(conditional) probability $| \lb f, a_i \rb |^2$. 
The total probability that the postselection succeeds is 
$$
\sum_{i=1}^2 |\lb a_i, s \rb |^2 | \lb f, a_i \rb |^2
\q.
$$
Hence the conditional expectation of $A$ given that the postselection
succeeds is
\beq
\lbl{eq120}
\frac
{\alpha_1 |\lb a_1, s \rb |^2 |\lb f, a_1 \rb |^2  
+ \alpha_2 |\lb a_2, s \rb |^2 | \lb f, a_2 \rb |^2 
}
{  |\lb a_1, s \rb |^2 | \lb f, a_1 \rb|^2  
+  |\lb a_2, s \rb |^2 | \lb f, a_2 \rb|^2
}
\q.
\eeq
It is easy to see that this need not equal \re{eq110}, 
$\Re (\lb f, As \rb / \lb f, s \rb)$.  For example, \re{eq120} is a convex
linear combination of $\alpha_1, \alpha_2$, and therefore cannot be arbitrarily
large, whereas 
$\Re (\lb f, As \rb / \lb f, s \rb)$ can be arbitrarily large when 
$f$ is nearly orthogonal to $s$ and the numerator $\lb f, As \rb$ is not
close to 0.   (Such examples are easy 
to construct). 
 
Therefore, if sentence 2 is to be true, the ``average value of $A$ conditioned
on postselection to $f$'' cannot refer to normal measurement in $S$.  To what
could it refer?  If it refers to measurement in $M$ of the normalized 
average value
of $B$ postselected to $f$, then sentence 2 becomes a tautology, true
by definition and containing no useful information..   
\subsection{``Weak values'' are not unique}

We think that the strongest argument that the ``weak value'' 
$\Re (\lb f, As \rb / \lb f, s, \rb )$ does not correspond to any
simple physical attribute of the system $S$ 
which is being weakly measured is that one can obtain 
{\em different} expressions
corresponding to weak measurements in $M$ followed by postselection to $f$.  
These different expressions will be obtained by reasoning conceptually
identical to the reasoning which led to \re{eq110}.  

Consider the setup described above of a system $S$ coupled to 
a two-dimensional meter system $M$ with orthonormal basis $m_0, m_1$.  
Given a Hermitian operator $A$ on $S$ and an initial state $s$ of $S$,
we want to approximate $\lb s, As \rb$ to arbitrary accuracy by measuring
the average value of an observable $B$ on $M$.  The notation will be the
same as in the previous discussion leading to the ``weak value'' \re{eq110}. 

Let $V$ be any unitary operator on $S$ satisfying $Vs = s$.
In place of the state $\tilde{r}(\epsilon)$ of $S \tnr M$ in \re{eq105}, 
define 
\begin{eqnarray}
\lbl{eq130}
\rhat(\epsilon) &:=&  
  (V \tnr I)\tilde{r}(\epsilon) \nonumber\\ 
&=& (V\tnr I) \frac{s \tnr m_0 \sqrt{1-\epsilon^2} + As \tnr m_1 \epsilon } 
{\sqrt{1 - \epsilon^2 + \epsilon^2 |As|^2}} \nonumber \\
&=&  \frac{Vs \tnr m_0 \sqrt{1-\epsilon^2} + VAs \tnr m_1 \epsilon } 
{\sqrt{1 - \epsilon^2 + \epsilon^2 |As|^2}} \nonumber\\ 
&=& \frac{s \tnr m_0 \sqrt{1-\epsilon^2} + VAs \tnr m_1 \epsilon } 
{\sqrt{1 - \epsilon^2 + \epsilon^2 |As|^2}} 
\q.
\end{eqnarray}
The condition $Vs = s$ guarantees that the 
corresponding state of $S$ will approximate $s$ for small $\epsilon$. 

Define the operator $B$ on $M$ as in \re{eq90} with $B_{00} = 0$
and $B_{01} = 1/2 = B_{10}$.
Since $V \tnr I$ and $I \tnr B$ commute,
$$ 
\lb \rhat,  (I \tnr B) \rhat \rb =
\lb (V \tnr I) \tilde{r},  (I \tnr B) (V \tnr I) \tilde{r} \rb = 
\lb  \tilde{r},  (I \tnr B)  \tilde{r} \rb \approx \epsilon \lb s, As \rb, 
$$
and
\beq
\lbl{eq142}
\lim_{\epsilon \mapsto 0} 
\frac{\lb \rhat(\epsilon), (I \tnr B) \rhat(\epsilon) \rb}{\epsilon} = \lb s , As \rb
\q,
\eeq
so averaging measured values of $I \tnr B$ in state $\hat{r}$ and normalizing
by dividing by $\epsilon$ 
is a weak measurement protocol
in the sense of Definition \ref{def1}.

As in the discussion leading to \re{eq110}, suppose that after measuring
$I \tnr B$ with $S \tnr M$ in state $\rhat (\epsilon)$, 
we postselect the state of $S$ to $f \in S$ (which  as previously explained
means a successful measurement of $P_f \tnr I$ in $S \tnr M$). As before, 
the conditional expectation $E_\epsilon(B|f)$ of $B$ given success 
of the postselection is 
\begin{eqnarray}
\lbl{eq150} 
E_\epsilon(B|f) &:=& 
\frac{\lb \rhat(\epsilon), (P_f \tnr B) \rhat (\epsilon) \rb}
{\lb \rhat, (P_f \tnr I) \rhat \rb} \\
&=& 
\frac{ (1 - \epsilon^2) B_{00}\lb s, P_f s \rb   + 
2\epsilon \sqrt{1-\epsilon^2} \Re (B_{10} \lb s, P_f VAs \rb) + 
\epsilon^2 B_{11}\lb VAs, P_f VAs \rb}  
{ (1 -\epsilon^2) \lb s, P_f s \rb + \epsilon^2 \lb VAs, P_f VAS \rb} 
.  \nonumber
\end{eqnarray}
Although we are assuming that $B_{00}= 0 $ and $B_{10} = 1/2 = B_{01}$, 
we included these above so that $\lb \rhat, (P_f \tnr I) \rhat \rb$ in the
denominator could be read off by setting $B := I$ in the numerator.  Also
note that the normalization factor making $|\rhat| = 1$ is the same in the
numerator and denominator and hence cancels. 

Under our assumption $B_{00} = 0$, 
the expression \re{eq150} for $E_\epsilon$ is of order $\epsilon$. 
As in the previous protocol leading to 
the traditional ``weak value'' \re{eq110}, we normalize 
by dividing by $\epsilon$ and take the limit as $\epsilon \goesto 0$,  
obtaining as the result of our new protocol the new ``weak value'' 
for $\lb s, As \rb$ conditional on postselection to $q$:
\beq
\lbl{eq160}
\lim_{\epsilon \goesto 0} \frac{E_\epsilon (B|f)}{\epsilon} = 
\frac{\Re \lb s, P_f V As \rb}{\lb s, P_f s \rb}
= \Re \frac{\lb f, VAs \rb} {\lb f, s \rb}
,
\eeq 
where the last inequality was obtained by manipulations similar to 
those leading to \re{eq110},to which this this reduces when $V = I$. 

To see that \re{eq110} and \re{eq160} need not be equal, take $S$
to be two-dimensional with orthonormal basis $s, s^\perp$.  Define $A$
and $V$
by the following matrices with respect to this basis,
$$
A :=
\left[
\begin{array}{ll}
0 & 1 \\
1 & 0
\end{array},
\right]
\q
V :=
\left[
\begin{array}{ll}
1 & 0 \\
0 & \eta
\end{array},
\right]
$$ 
with $|\eta| = 1,$
so that $Vs = s$, $Vs^\perp = \eta s^\perp$, $As = s^\perp$ and $As^\perp = s$. 
Let $f := (s + s^\perp)/\sqrt{2}$. Then the numerator of \re{eq160} becomes
$$
\Re \lb s, P_f VAs \rb = \Re (\lb s,  P_f s^\perp \rb \eta)
 = \Re(\lb s ,  \lb f, s^\perp \rb f  \rb \eta) = \Re \eta/2.   
$$
Equation \re{eq160} for this special case becomes 
\beq
\lbl{eq170}
\lim_{\epsilon \goesto 0} \frac{E_\epsilon (B|f)}{\epsilon} = 
\Re \eta
\q.  
\eeq 
Obviously, this is not independent of $\eta$, as it would have to be 
if ``weak values'' were unique.  
Any real number between $-1$ and 1
can be obtained as a ``weak value'' for the specified  $s, f$ and $A$. 

The title of this subsection, ``weak values are not unique'', summarizes its 
conclusion.  The reasoning leading this conclusion may be summarized 
as follows.  Given an observable $A$ on $S$ and a reproducible 
pure state $s$ of $S$, 
we defined a one-parameter family, indexed by $|\eta | = 1$, 
of weak measurement protocols (in the sense of Definition 1) to approximate
$\lb s, As \rb$ while negligibly changing the state $s$ of $S$.
Applying one of these weak measurement protocols conditional on successful
postselection of the (negligibly changed) state of $S$ to $f$ gives a 
conditional expectation called a ``weak value'' of $A$.  This ``weak value''
is not independent of the protocol, i.e., not independent of $\eta$.  

Since all these different ``weak values'' were obtained by conceptually
identical reasoning, we see no reason to identify any one of them with
some intrinsic, measurement-independent property of system $S$.  The ``standard''
weak value \re{eq110} is one of these and seems generally identified in the 
literature (often implicitly) with the expectation of $A$ conditional 
on successful postselection to $f$.  We think such an identification fallacious.
\section{Historical summary and simplification of traditional approach}
\subsection{Overview}
This section describes an approach similar, but not identical, to that 
pioneered by Aharonov, Albert, and Vaidman 
\cite{AAV}, which builds on a classical theory of measurement due to 
von Neumann \cite{vN}.   
The cited paper of Aharanov, Albert, and Vaidman will be called ``AAV'' below.  
It culminates in a formula for weak values identical to \re{eq160} 
except that the real part symbol $\Re$ is omitted.

The omission is rather curious.  Their footnote 4 seems to recognize
that the real part should be taken, but the uncontrolled (and questionable) 
approximations in the main text yield \re{eq160} without the real part.  
Subsequent literature
mostly takes the real part, so when we refer to the ``usual'' weak value formula,
we mean (expressed in the notation of \cite{AAV}, 
which is commonly used in the literature)
\begin{eqnarray}
\lbl{eq180}
\lefteqn{\mbox{``usual'' weak value of observable $A$}}\nonumber\\
& &\mbox{ with initial state $\psi_{in}$ 
and post-selected to state $\psi_f$}  \nonumber \\
&=& 
\Re \frac{\lb \psi_f, A \psi_{in} \rb}{\lb \psi_f, \psi_{in}\rb}
.
\end{eqnarray}
In notation used above and to be used below, this would read
\begin{eqnarray}
\lbl{eq181}
\lefteqn{\mbox{``usual'' weak value of observable $A$}}\nonumber\\
& &\mbox{ with initial state $s$ 
and post-selected to state $f$}  \nonumber \\
&=& 
\Re \frac{\lb f, A s\rb}{\lb f, s\rb}
.
\end{eqnarray}

All literature on weak values known to me gives the impression that \re{eq180}
(or \re{eq180} without the real part)
is a universal formula which is to be expected in all experimental
situations.  We have seen in the last section that this is not so. 

Most ``derivations'' 
(which really should be called ``motivations'') of \re{eq180}
in the literature are mathematically imprecise and overly complicated.  
We hope that a  
cleaner motivation 
may help clarify the domain of applicability of \re{eq180}.
Our conclusion will be that \re{eq180} may well hold in some experimental
situations, but that claims to its universality should be critically examined.
\subsection{Preparation of the initial state for a weak measurement}
The analysis of the previous section started with initial states called 
$\tilde{r}(\epsilon)$ in \re{eq105} or $\hat{r}(\epsilon)$ in \re{eq130}, and 
it was implicitly assumed that such states are physically realizable.
By contrast, the literature such as AAV  generally starts with an
initial 
product state  $s \tnr m \in S \tnr M$ 
(product states are generally considered physically 
realizable) and obtains the desired entangled state to measure in 
the meter system by applying 
a unitary operator $U$ to obtain  $ U(s \tnr m)$ as the analog of our 
slightly entangled starting state $\tilde{r}(\epsilon)$.  The unitary operator
$U$ is usually considered as a time evolution operator $U = e^{-iHt}$ with 
$H$ the Hamiltonian, and its only function is to assure that  
the initial state $U(s \tnr m)$ is physically realizable.  
In our version of the AAV approach, $t$ will be considered as a small 
parameter, which subsequently will be called $\epsilon$.

Both von Neumann and AAV use a ``meter'' Hilbert space $M$ defined as 
the space of all square integrable complex-valued functions on the real
line $\bR$.  This space is known to physicists as the Hilbert space of
a single spinless particle in one dimension 
and to mathematicians as $L^2(\bR)$,
the space of all complex-valued functions $g = g(q)$ of a real variable
$q$ which are square-integrable:
$$
\int_{- \infty}^\infty \, |g(q)|^2 \, dq < \infty \q.
$$
The inner product on $M = L^2(\bR)$ is defined as usual for $g, h \in M$ by
$$
\lb g, h \rb := \int_{- \infty}^\infty\, g(q)^* h(q) \, dq \q.  
$$
Two important Hermitian operators on $L^2(\bR)$ are the position operator
$Q$ defined for $g \in L^2(\bR)$ satisfying certain technical
conditions (which we do not list here because they will not be 
important to us) by 
\beq
\lbl{eq183} 
Qg(q) := q g(q)\q\mbox{for all $ q \in \bR$},
\eeq
and the momentum operator $P$ defined by 
\beq
\lbl{eq184}
(Pg)(q) := -i \frac{dg}{dq} \q \mbox{for all $q \in \bR$}. 
\eeq

For any real $\alpha$, if 
we expand $e^{-i\alpha P}$ in a formal power series and apply that to
a smooth function $g \in L^2(\bR)$, using Taylor's theorem we obtain   
\beq
\lbl{eq185}
(e^{-i\alpha P} g)(q) = g(q - \alpha) \q.
\eeq
In other words, $e^{-i\alpha P} $ translates the graph of $g$ a distance of
$\alpha $ units to the right.  We shall denote this translate 
of $g$ by $g_{\alpha}$:  
\beq
\lbl{eq186}
g_\alpha (q) := g(q - \alpha) 
\q.
\eeq  
This calculation was purely formal,
but it is well known how to formulate definitions under which
it can be rigorously proved.%
\footnote{In careful treatments of the foundations of quantum mechanics such
as \cite{Mackey}, the momentum operator $P$ is {\em defined} as the infinitesimal
generator of translations, so that $(e^{-i\beta P}f)(q) = f(q - \beta)$ is true
by definition.  Then it is later verified that $P = -i d/dq$ on an
appropriate domain.}

In the finite dimensional context considered in the last section, physical
realizablility of the entangled state to be meter measured 
($\hat {r}$ or $\tilde{r}$ of the previous sections)
is not an issue because 
any unitary operator on a finite dimensional Hilbert space can be physically
realized to an arbitrary approximation by physically constructible quantum
gates (\cite{N/C}, Chapter 4), and any state can obviously be obtained
by applying a unitary operator to a product state.  

The issue in infinite-dimensional contexts is sidestepped 
by both von Neumann and AAV via the assumption
that the required Hamiltonian will be physically realizable.  
They both
essentially use a Hamiltonian $H$ of the form 
\beq
\lbl{eq190}
H = A\tnr P \q,
\eeq
where $A$ is the Hermitian operator on $S$ whose expectation 
$\lb s , A s \rb$ is
to be approximated by a meter measurement. 
\footnote{
AAV uses a slightly more complicated expression, but in the context of their
argument it is essentially the same as \re{eq190}.  
}

\section{AAV's extension of Von Neumann's theory of measurement} 
\subsection{Our adaptation of Von Neumann's general framework}

Von Neumann \cite{vN} did not consider ``weak'' measurements 
{\em per se}. For 
the purposes of dealing with them, we shall modify \re{eq190}
by inserting
a small positive parameter $\epsilon$ to measure the strength of the 
interaction:
\beq
\lbl{eq200}   
H(\epsilon) := \epsilon A \tnr P \q.
\eeq
(Alternatively, $\epsilon$ could be considered as a small time; the mathematics
is insensitive to this sort of variation of the physical picture.)
Eventually, we shall take a limit $\epsilon \goesto 0$ after appropriate
normalization.

The operator to be measured in the meter system (the analog of the operator 
$B$ of the previous sections) will be the position operator $Q$.
We will see that measuring $Q$ in the state $e^{-iH(\epsilon)} (s \tnr m)$
will yield an average value of $\epsilon \lb s, As \rb$.
 
The setup of AAV is superficially different in that
they interchange $Q$ and $P$, so that their ``meter'' measures
the momentum of the ``pointer'' rather than its position.  Since Fourier
transformation implements a unitary equivalence which takes  $P$  to $-Q$
and $Q$ to $P$, 
there is no essential mathematical difference between  the two formulations. 

The significance of the Hamiltonian \re{eq200} is most easily seen in 
the case in which $A$ has only one eigenvalue $\alpha$, so that 
$A = \alpha I$, where $I$ is the identity operator on an $n$-dimensional
Hilbert space.  Let $a_1, a_2, \ldots , a_n$ be an orthonormal basis 
for $S$.  Then $S \tnr M$ identifies naturally with a direct sum
of $n$ copies $[a_i] \tnr M \cong M $ of $M$, where $[a_i]$ denotes the 
one-dimensional subspace of $S$ spanned by $a_i$, and $\cong$ denotes
isomorphism.  Moreover, each of the
copies $[a_i] \tnr M$ is invariant under $H(\epsilon)$ and may 
be naturally identified with $M = L^2(\bR)$.  When so identified,  
$e^{-iH(\epsilon)}$ acts 
as translation by $\epsilon \alpha$:  
$(e^{-iH(\epsilon)}g)(q) = g(q - \epsilon \alpha)$. 

Thus when $A$ has just one eigenvalue $\alpha$, 
$e^{-iH \epsilon \alpha}$  simply translates the initial probability
distribution of pointer readings by $\epsilon \alpha$.  If the starting 
meter state $m$ yields an average reading $\gamma = \lb m, Qm \rb$, 
then
the average reading in a state $e^{-iH(\epsilon)}(s \tnr m)$ will
be $\gamma + \epsilon \alpha$.  For simplicity of language, we choose 
the origin of $\bR$ so that 
$\gamma =0$, i.e., so that the average
meter reading in state $m$ is 0. 

We ``read the meter'' by starting with the reproducible 
product state $s \tnr m$, applying $e^{-iH(\epsilon)}$, and measuring 
$I\tnr Q$.  Repeating this many times  obtains an average value for 
$I \tnr Q$ of $\epsilon \alpha$, which for fixed $\epsilon$ can be determined
to arbitrary accuracy by measuring sufficiently many times. 
Dividing by $\epsilon$ gives $\alpha$.  It seems reasonable to hope that 
this measurement might be ``weak'' in the sense of Definition
\ref{def1} because for small $\epsilon$, $e^{-iH(\epsilon)}
(s \tnr m) = s \tnr e^{-i\epsilon \alpha P} m = s \tnr m_{\epsilon \alpha}$ 
is arbitrarily close
in norm to $s_0 \tnr m_0$ (because translations are strongly continuous).  
\footnote{
That is a hand-waving plausibility argument of the type often accepted 
in the physics literature, not a proof.  It is actually wrong in the sense
that it cannot easily be made into a proof; one sticking point is that 
the trace and operator norms are not equivalent when $M$ 
is infinite-dimensional.  However, the weakness of the measurement can 
be justified by other means under additional hypotheses.  Appendix 2 
examines the surprisingly delicate issue of ``weakness'' in more detail.
}

The situation is similar if $A$ has several eigenvalues $\alpha_1,
\alpha_2, \ldots , \alpha_n$ (not necessarily distinct) 
with a corresponding  orthonormal basis $a_1, a_2, \ldots, a_n$ of 
eigenvectors:  $A_i a_i = \alpha_i a_i, i = 1, 2, \ldots , n $.
 In that 
case, $S \tnr M$ decomposes as a direct sum $\oplus_i ( [a_i] \tnr M)$, where 
each $[a_i] \tnr M$ is invariant under $H(\epsilon)$ and has a natural 
indentification with  $M = L^2(\bR)$.
The action of $e^{-iH(\epsilon)}$ on 
$[a_i] \tnr M$ is as described above when $A$ had only one eigenvalue.

Denote the decomposition of the initial normalized state $s \in S$ as a direct
sum of orthogonal states in $[a_i]$ as
\beq
\lbl{eq205}
s = \oplus_i \ s_{i}\q, 
\eeq
with $s_{i} \in  [a_i] $ (so that each $s_i$ is a multiple of $a_i$), and 
$\sum_i |s_i|^2 = 1$. 
Then 
$$\lb s \tnr m, (A \tnr I)(s \tnr m) \rb =
\sum_i \alpha_i |s_i|^2
$$ 
is the expectation $\lb s, A s \rb$ 
of $A$ in the state $s$, which we shall show is also equal to
\beq
\lbl{eq215}
\frac{\lb e^{-iH(\epsilon)}(s \tnr m), Q e^{-iH(\epsilon)}
(s \tnr m) \rb}
{\epsilon} \q.
\eeq
Denoting (as always) by  $g_{\beta}$ the 
translate of a function $g$ by $\beta$, $g_\beta (q) := g(q - \beta)$, 
 we have 

\begin{eqnarray}
\lbl{eq220}
\lb e^{-iH(\epsilon)}(s \tnr m), (I \tnr Q)e^{-iH(\epsilon)} (s \tnr m) \rb
&=&  
\lb \oplus_j s_{j} \tnr (m)_{\epsilon \alpha_j}, 
\oplus_k
 s_{k} \tnr Q((m)_{\epsilon\alpha_k}) \rb \nonumber \\
&=&\sum_j |s_j|^2 \lb {m}_{\epsilon \alpha_j}, Q {m}_{\epsilon \alpha_j} \rb
\nonumber \\
&=& \sum_j  |s_j|^2 \int_{- \infty}^\infty dq \,  
m^*(q - \epsilon \alpha_j)\, q \,  
m(q - \epsilon \alpha_j) \nonumber \\
&=& \sum_j  |s_j|^2 \int_{-\infty}^\infty dq\, m^* (q) 
(q + \epsilon \alpha_j)m(q) \nonumber \\
&=& \sum_j   |s_j|^2 \epsilon \alpha_j \nonumber \\
&=& \epsilon \lb s, As \rb \q.
\end{eqnarray}
In the second line we used the orthogonality of the the $s_j$ to 
convert a $\sum_{j,k}$ into a $\sum_j$, and the passage to the next to last
line uses the assumption that the expectation $\lb m_, Qm \rb$
 of $Q$ in the state $m$ is 0.

The expectation $\lb s, As \rb$ 
of $A$ can be determined to arbitrary accuracy by averaging a sufficiently 
large number of measurements of pointer position $Q$ 
for small, fixed $\epsilon$, and finally dividing
by $\epsilon$:
\beq
\lbl{eq230}
\lb s, As \rb = 
\frac{\lb 
e^{-iH(\epsilon) s}(s \tnr m), 
(I \tnr Q)e^{-iH(\epsilon) s}(s \tnr m) \rb}
{\epsilon} 
\q.
\eeq
The normalized (by dividing by $\epsilon$, which is our definition of 
``normalized'' in this context)
conditional expectation of $(I \tnr Q)$ given postselection to $f \in S$
is 
\begin{eqnarray} \lbl{eq240}
\frac{1}{\epsilon}
\frac{\lb 
e^{-iH(\epsilon)}(s \tnr m), (P_{f} \tnr Q)e^{-iH(\epsilon)} 
(s \tnr m) \rb
}
{
\lb e^{-iH(\epsilon)}(s \tnr m), (P_{f} \tnr I)
e^{-iH(\epsilon)}(s \tnr m) \rb 
}
\q.
\end{eqnarray}
The numerator  is the expectation in the state 
$e^{-iH(\epsilon)}(s \tnr m)$ of the product of the commuting observables 
$I \tnr Q$ and $P_{f} \tnr I$. 
This product has the value of $Q$ if the postselection succeeds
and 0 otherwise.  The denominator is the probability that the postselection
succeeds.  

Assuming weakness of the measurement, i.e., that
after applying $e^{-i H(\epsilon)}$ to $s \tnr m$ and then measuring $Q$,
the  state of $S$ will be close to the original state $s$,
we have just described a weak measurement protocol (in the sense of 
Definition \ref{def1}) 
for approximating $\lb s, As \rb$ to arbitrary accuracy while
making an arbitrarily small change in the original state $s$ of $S$.
The technical problem of proving weakness will be examined in Appendix 2. 
For the moment, we assume it.   

The end of Section 4 pointed out the logical fallacy of identifying \re{eq240}
with the expectation of $ A$ in the initial state $s$
conditional on postselection to $f$, as equation \re{eq230} tempts.  
If nevertheless we
make this identification (as AAV do), 
then we shall show that 
the limit as $\epsilon \goesto 0$ of \re{eq240} 
produces a ``weak value'' 
\beq
\lbl{eq242}
\Re
\frac{\lb f, As\rb}
{\lb f, s \rb}   
\eeq
for $A$.   
This coincides with the ``weak value'' obtained by 
Lundeen and Steinberg \cite{LS} by different methods.  By contrast, the 
``weak value'' obtained by AAV was
\beq
\lbl{eq245}
\frac{\lb f, As\rb}{\lb f, s \rb}   
\q.
\eeq

Before calculating \re{eq240}, we continue the historical exposition
by indicating how AAV obtained \re{eq245}.  
Instead of assuring weakness
of the interaction by making the Hamiltonian $H(\epsilon)$ close to 0, as
we did above, they use the above Hamiltonian 
$H(\epsilon)$ with $\epsilon := 1$ .  To obtain a ``weak'' interaction
which affects the state of $S$ only slightly, they  
vary the initial meter state $m$, which they assume
real with a square (which is a probability distribution on $\bR$) of Gaussian form:
\beq
\lbl{eq250}
m^2(q) = \frac{\exp(- q^2/2\sigma^2)}{(2 \pi)^{1/2} \sigma}
\q. 
\eeq  
They attempt to obtain weakness of the measurement by taking $\sigma$ large, 
meaning that the Gaussian is very spread out.
One hopes that this might  assure weakness because then $(m)_{\alpha_i}$ 
will be
close in norm to $m$ for all of the eigenvalues $\alpha_i$ of $A$, so that
tracing the state $e^{-iH}(s \tnr m)$ over $M$ to obtain the  state of $S$
might be hoped to yield
\beq
\lbl{eq255}
e^{-iH}(s \tnr m) = \sum_i  s_{i} \tnr m_{\alpha_i} 
\approx \sum_i  s_{i} \tnr m 
\stackrel{\mbox{trace}_M}{\rightarrow} \sum_i \ s_{i} = s
.  
\eeq 
(The above implicitly identifies vector states $v \in S \tnr M$ 
with the corresponding density matrices $P_v$, in order to apply $\trM$.)%
\footnote{This argument is highly suspect because the ``$\approx$'' 
is only obvious for the Hilbert space norm, and $\trM$ is not continuous
from the Hilbert space norm on $S \tnr M$ to any norm on $S$.  However, Appendix 2 
shows how it can be fixed.  We include it
primarily to provide a motivation for the AAV approach which would probably
be accepted in the physics literature and secondarily to underscore the
need for care when making approximations. 
}

Their argument involves detailed calculations with Gaussians employing 
uncontrolled approximations.%
\footnote{A. Peres \cite{Perescom} has characterized equation (3) of AAV
as a ``faulty approximation''.  
It seems surprising that questions about basic mathematical procedures leading 
to new and striking conclusions have remained unresolved in the literature
for over 20 years.} 
After calculating \re{eq240}, we shall note that the same argument 
(assuming weakness, as do AAV)
is easily adapted to produce a mathematically rigorous version of the 
calculation attempted by AAV, but concluding with a different result. 

\subsection{Calculation of the AAV-type ``weak value''}

Before beginning, we summarize the notation to be used. 
Since we shall be working with a ``meter'' Hilbert space $M := L^2(\bR)$,
which is a function space,
we change notation for the norm of an element $g$ of $M$ to 
$$||g|| := \left[ \int_{\bR}  |g(q)|^2 \, dq\, \right]^{1/2} \q,$$  
to distinguish it from its absolute value $|g(q)|$.
The Hilbert space $S$ of the system of primary interest will be assumed
finite dimensional, and we continue to denote the norm of elements of 
$s$ as $|s|$.  Also, we denote the norm of $u \in S \tnr M$ by $|u|$.
 
We assume given a reproducible product state 
$s \tnr m \in S \tnr M = S \tnr L^2(\bR)$ with $|s| = 1 = ||m||$ 
to which we shall
apply the unitary operator $e^{-iH(\epsilon)} = e^{-i \epsilon A \tnr P}$ to
obtain a state on which to perform a meter measurement, which means measuring
$(I \tnr Q)$.  After performing the meter measurement, we postselect
to a given final state $f \in S$.  Postselecting to $f$
means measuring $P_f \tnr I$%
\footnote{More precisely, we model the procedure of measuring $Q$ followed
by postselection to $f$ as a measurement of $P_f \tnr Q$, the value of which
is the value of $Q$ if $f$ is obtained, and 0 otherwise.  Discarding the
results for which $f$ is not obtained then corresponds to division of the 
average value of this measurement by the average value of a measurement
of $P_f \tnr I$. 
But it is easier to think and speak of first measuring $Q$, 
then $P_f$.  The subtle difficulty in the latter way of thinking is that
the state of the system after a measurement of $Q$ 
(but before the postselection) is not precisely defined by standard quantum
mechanics, as discussed in Appendix 2. 
}
and discarding results in which 
this last measurement gives 0.
The results which are not discarded 
are then averaged and normalized (by dividing by $\epsilon$) 
to produce a final ``weak value'' which is often interpreted
in the literature (incorrectly, in our view) as an approximation
(becoming exact as $\epsilon \goesto 0$) to the average value of $A$
in the state $s$ conditioned on successful postselection to $f$.

We assume that $S$ is finite dimensional.  Let $A: S \goesto S$ 
be a given Hermitian operator, 
for which we desire that the average meter measurement will
approximate $\lb s, As \rb$.  
Our goal will be to calculate
\begin{eqnarray} 
	\lbl{eq400}
	\lim_{\epsilon \goesto 0}
	\frac{1} {\epsilon}
	\frac{\lb 
	e^{-iH(\epsilon)}(s \tnr m), (P_f \tnr Q)e^{-iH(\epsilon)} 
	(s \tnr m) \rb
	}
	{\lb e^{-iH(\epsilon)}(s \tnr m), (P_f \tnr I)
	e^{-iH(\epsilon)(s \tnr m)} \rb 
	}
	\q,
	\end{eqnarray}
which corresponds to the procedure just described.  The numerator of the 
second fraction is the expectation of the measurement of the product of the 
commuting operators $I \tnr Q$ and $P_f \tnr I$, which is the expectation of 
a measurement of $Q$ with measurements in which the postselection fails 
counted as zeros.  The denominator is the probability that the postselection
succeeds, and the entire quotient the expectation of $Q$ conditional on success
of the postselection.

	The result of the calculation will be the ``weak value'' 
	\beq
	\lbl{eq410}
	\Re 
	\frac{\lb f, As \rb}
	{\lb f, s \rb}
	\q, 
	\eeq
	assuming that $\lb f, s \rb \neq 0$.
	When $\lb f, s \rb = 0$, the expression is undefined.

However to obtain this result, we shall need the following additional 
assumptions on the initial meter state $m \in L^2(\bR)$.
All but one are mild regularity and growth
conditions.  All are satisfied by the Gaussian meter states
used by AAV.  These assumptions are:
\begin{enumerate}
\item  We assume that $m$ is a real-valued function. 
A previous version of this paper stated that this was not an essential 
assumption, but that was a mistake.  
The proof as given in the rest of the section does require this assumption.
A later section added after the mistake was discovered completes the proof
to be given below without this assumption, obtaining still more 
non-traditional ``weak values''. 
\item
We continue to assume, as in the preceding discussion, 
that $\lb m , Qm \rb = 0$, 
i.e., that the average meter reading is initially  ``$0$''. 
\item  We asume that $m$ satisfies the growth condition 
$$
\lim_{q \goesto \pm \infty} = qm^2(q) = 0
$$
\item
We assume that $m$ has a continuous second derivative 
$m^{\prime\prime} = m^{\prime\prime} (q)$
satisfying the growth condition  
$$
\lim_{q\goesto \pm \infty} q^3 |m^{\prime\prime}(q)| = 0 \q.
$$
\end{enumerate}
These assumptions can be weakened in various ways which seem not very
interesting.  Most are the kind of regularity and growth conditions
typically assumed without mention in physics calculations. 
The statements given were chosen for their simplicity.

Actually, we shall perform a more general calculation which will yield
a ``weak value'' which is in general nothing like the traditional
value \re{eq242}, in order to illustrate
that even in the AAV context, ``weak values'' are not unique.
The ideas are the same as in the finite dimensional 
example of nonunique weak values in Section 5.4.%
\footnote{
That section employed a unitary operator $V$ on $S$ with $Vs = s$.
The choice $V :=I$ yielded the traditional weak value 
$\Re (\lb f, As \rb/ \lb f, s \rb$, and other choices yieled other
weak values.  By using a non-real meter function $m \in L^2(\bR)$, one
can obtain non-traditional weak values even with $V := I$, as explained 
in a later section.  Thus the reader
should feel free to set $V := I$ on first reading.
}

We assume given a normalized ``initial state'' $s$ of $S$, 
and a normalized ``meter state'' $m$ of $M$
satisfying the above conditions.
As before, $A$ is a given Hermitian operator on $S$ whose expectation in
the state $s$, $\lb s, As \rb$ is to be measured. Let $a_1, a_2, \ldots, 
a_n$ an orthonormal basis of eigenvectors for $A$, and write  
\beq
\lbl{eq502}
s = \sum_{i=1}^n s_i 
,\q
\mbox{with $s_i$ a multiple of $ a_i$ and } \sum_i |s_i|^2 = 1. 
\eeq

For the more
general calculation, we also assume given a unitary operator $V: S \rightarrow S$
which satisfies $Vs = s$. 
In place of the unitary operator $e^{-i H(\epsilon )}$ which 
prepared the
	 state for the meter measurement  (by applying it to $s \tnr m$), 
we shall use
	\beq
	\lbl{eq420}
	(V\tnr I) e^{-i H(\epsilon )} = (V \tnr I)e^{-i\epsilon A \tnr P} 
	\q, 
	\eeq
	resulting in an initial state
	\beq
	\lbl{eq430}
	(V\tnr I) e^{-i H(\epsilon )} (s \tnr m) = 
\sum_j Vs_j \tnr m_{\epsilon \alpha_j} 
\eeq
just before the meter measurement, which will use
\beq
\lbl{eq435} 
\frac{ \lb (V\tnr I) e^{-i H(\epsilon )} s \tnr m, (I \tnr Q) 
(V \tnr I) e^{-i H(\epsilon )} s \tnr m  \rb}
{\epsilon}
\eeq
to approximate $ \lb s, As \rb$.
The ``weak value'' to be calculated is \re{eq400} with 
$e^{-iH(\epsilon)}$ replaced by $(V\tnr I) e^{-iH(\epsilon)}$:
\beq
\lbl{eq520}
	\lim_{\epsilon \goesto 0}
\frac{1}{\epsilon}
\frac{\lb 
(V\tnr I)e^{-iH(\epsilon)}(s \tnr m), (P_{f} \tnr Q)(V\tnr I)e^{-iH(\epsilon)} 
	(s \tnr m) \rb
	}
	{
	\lb (V \tnr I) e^{-iH(\epsilon)}(s \tnr m), (P_f \tnr I)
(V\tnr I)e^{-iH(\epsilon)}(s \tnr m) \rb 
	}
	\q.
\eeq
Of course, taking $V := I$ recovers the AAV-type situation 
previously discussed and \re{eq400}. 
The only reason for the assumption $Vs = s$ is to assure the weakness
of the measurement, as will be shown in Appendix 2.  Since this assumption will
play no role in the following calculation, we refrain from replacing
$Vs$ by $s$.

First we calculate the limit as $\epsilon \goesto 0$ of the denominator
of \re{eq400}.
\begin{eqnarray}
\lbl{eq530}
\lefteqn{\lim_{\epsilon \goesto 0} 
\lb (V\tnr I) e^{-iH(\epsilon)}(s \tnr m), (P_f \tnr I)(V\tnr I)
e^{-iH(\epsilon)} 
	(s \tnr m) \rb} \nonumber \\
&=& 
\lim_{\epsilon \goesto 0} \lb (V\tnr I) \sum_i s_i \tnr m_{\epsilon \alpha_i} ,
 (P_f \tnr I) (V \tnr I) \sum_j s_j \tnr m_{\epsilon \alpha_j} 
\rb \nonumber\\
&=&
\sum_{i,j}  
\lb Vs_i , P_f Vs_j \rb \lim_{\epsilon \goesto 0}
\lb m_{\epsilon \alpha_i} , m_{\epsilon \alpha_j} \rb \nonumber\\
&=&  
\sum_{i,j}  \lb Vs_i , P_f V s_j \rb \nonumber\\
&=& \lb f, Vs \rb \lb Vs , f \rb 
\q,
\end{eqnarray}
where we have used the fact that translations are continuous in the Hilbert
space norm on $L_2(\bR)$ to eliminate the factor involving the inner product
of the $m$'s, and obtained the last line by recalling 
from \re{eq502} that $\sum_i s_i = s$.
 
Next we calculate 
\beq
\lbl{eq535}
\lim_{\epsilon \goesto 0} 
\frac{ \lb (V\tnr I) e^{-i H(\epsilon)} s \tnr m, (T \tnr Q) 
V e^{-i H(\epsilon A)} s \tnr m  \rb}
{\epsilon}
,
\eeq
where $T: S \goesto S$ is an arbitrary Hermitian operator.  For $T := I$, this 
will validate \re{eq435} as an approximation to $\lb s, As \rb$, 
and for $T := P_f$, it will give the numerator of \re{eq520}.
We have    
\begin{eqnarray}
\lbl{eq540}
\lefteqn{\lb (V\tnr I) e^{-iH(\epsilon)}(s \tnr m), (T \tnr Q)(V\tnr I)
e^{-iH(\epsilon)} 
	(s \tnr m) \rb} \nonumber\\
&=& 
\lb (V\tnr I) \sum_i s_i \tnr m_{\epsilon \alpha_i} ,
 (TV\tnr Q) \sum_j s_j \tnr m_{\epsilon \alpha_j} 
\rb \nonumber\\
&=&
\sum_{i,j}  
\lb Vs_i , TVs_j \rb \lb m_{\epsilon \alpha_i} , (Qm)_{\epsilon \alpha_j} \rb 
\q.
\end{eqnarray}
The inner product involving $m$ is 
\begin{eqnarray}
\lbl{eq550}
\lb m_{\epsilon \alpha_i} , Qm_{\epsilon \alpha_j} \rb 
&=& \int_{q = -\infty}^\infty \, 
m^*(q - \epsilon \alpha_i)\, q \, m(q - \epsilon \alpha_j) \, dq
\nonumber\\
&=& 
\int \, m(q ) (q + \epsilon \alpha_i) 
m(q - \epsilon (\alpha_j - \alpha_i) )\, dq
\nonumber\\ 
&=& 
\int \, m(q)\epsilon \alpha_i m(q- \epsilon (\alpha_j - \alpha_i)) \, dq
\nonumber\\
&& +    
\int \, m(q)\, q \, [m(q) - m^\prime (q) \epsilon(\alpha_j - \alpha_i)] 
\, dq  +  O(\epsilon^2)
.
\end{eqnarray}
The second line was obtained by a linear change of variable in the integral. 
For the last line, we performed a power series expansion about $q$ 
to order 1 with remainder 
and used the growth conditions to estimate the remainder term
as $O(\epsilon^2)$.  In detail, for any $q$ and $\beta$,  
$$
m(q + \epsilon \beta) ) = m(q) + m^\prime (q) \epsilon \beta 
+ m^{\prime\prime}(\zeta_q)(\epsilon\beta)^2/2 
,
$$
where $\zeta_q$ is between $q$ and $q + \epsilon\beta$, and the growth
conditions on $m$ and $m^{\prime\prime}$ ensure that the integral of the
terms involving $m^{\prime\prime}$ is $O(\epsilon^2)$.

The first integral in the last line of \re{eq550}
is $\epsilon (\alpha_i + o(1))$ 
(because for any $\beta$, $||m_{\epsilon \beta}- m|| = o(1)$, where 
as usual in this context, $o(1)$ represents a term which goes to 0 as 
$\epsilon \goesto 0$. Integrating the second integral
 by parts yields
\begin{eqnarray}
\lbl{eq560} 
\lb m_{\epsilon \alpha_i} , Qm_{\epsilon \alpha_j} \rb 
&=& 
\epsilon [\alpha_i + \frac{\alpha_j - \alpha_i}{2}] + O(\epsilon^2) +  
 \epsilon o(1)
.
\end{eqnarray}
Hence 
\beq
\lbl{eq570}
\lim_{\epsilon \goesto 0} 
\frac{\lb m_{\epsilon \alpha_i} , Qm_{\epsilon \alpha_j} \rb } 
{\epsilon}
= \frac{\alpha_i + \alpha_j}{2}
\q.
\eeq 
Combining this with \re{eq540} gives \re{eq535} as 
\begin{eqnarray}
\lbl{eq580} 
\lefteqn{
\lim_{\epsilon \goesto 0} 
\frac{ \lb (V \tnr I) e^{-i H(\epsilon)} s \tnr m, (T \tnr Q) 
V e^{-i H(\epsilon A)} s \tnr m  \rb}
{\epsilon} 
}\nonumber\\
&=& \sum_i \sum_j \lb Vs_i, TVs_j \rb 
\frac{\alpha_i + \alpha_j}{2} 
 \nonumber \\
&=& \sum_i \frac{\lb Vs_i, TVAs \rb}{2} + 
\sum_j \frac{\lb VAs, TVs_j \rb}{2} 
\nonumber 
\\
&=& \frac{\lb Vs, TVAs \rb + \lb VAs, TVs \rb}{2} \nonumber\\ 
&=& \Re  \lb Vs, TVAs \rb 
\q.
\end{eqnarray}
To understand these manipulations, recall that the $s_i$ were defined
in \re{eq502} by $s = \sum_i s_i$ with $As_i = \alpha_i s_i$. 

Specializing \re{eq580} to $T:= I$ gives
\beq
\lbl{eq590} 
\lim_{\epsilon \goesto 0} 
\frac{ \lb (V\tnr I) e^{-i H(\epsilon)} s \tnr m, (I \tnr Q) 
(V\tnr I) e^{-i H(\epsilon A)} s \tnr m  \rb}
{\epsilon} 
= \Re  \lb Vs, VAs \rb =  \lb s, As \rb,
\eeq
justifying \re{eq435} as an approximation to $ \lb s, As \rb$. 

Specializing \re{eq580} to $ T := P_f$ gives
\beq
\lbl{eq600}
\lim_{\epsilon \goesto 0} 
\frac{ \lb (V\tnr I) e^{-i H(\epsilon)} s \tnr m, (P_f \tnr Q) 
(V \tnr I) e^{-i H(\epsilon A)} s \tnr m  \rb}
{\epsilon} 
= \Re (\lb f, VAs \rb \lb Vs, f \rb ) 
.
\eeq
Combining this with equations \re{eq530} and \re{eq580} yields the 
``weak value''
\begin{eqnarray}
\lbl{eq610}
\lefteqn{
\lim_{\epsilon \goesto 0}
\frac{\lb 
(V\tnr I)e^{-iH(\epsilon)}(s \tnr m), (P_{f} \tnr Q)(V\tnr I)e^{-iH(\epsilon)} 
	(s \tnr m) \rb
	}
	{\epsilon
	\lb (V \tnr I) e^{-iH(\epsilon)}(s \tnr m), (P_f \tnr I)
(V\tnr I)e^{-iH(\epsilon)}(s \tnr m) \rb 
	}
}\nonumber \\
&=& \frac{ \lb f, VAs \rb \lb Vs, f \rb + \lb VAs, f \rb \lb f, Vs \rb}
{2\lb f, Vs \rb \lb Vs, f \rb} 
\nonumber\\ 
&=& \Re \frac{\lb f, VAs \rb}{\lb f, Vs \rb } 
	\q.
\end{eqnarray}

Specializing to $V  = I$ yields the AAV-type ``weak value'':
\beq
\lbl{eq620}
\lim_{\epsilon \goesto 0}
\frac{\lb 
e^{-iH(\epsilon)}(s \tnr m), (P_{f} \tnr Q)e^{-iH(\epsilon)} 
	(s \tnr m) \rb
	}
	{\epsilon
	\lb  e^{-iH(\epsilon)}(s \tnr m), (P_f \tnr I)
e^{-iH(\epsilon)}(s \tnr m) \rb 
	}
 = \Re \frac{\lb f, As \rb}{\lb f, s \rb } 
\q	.
\eeq
%
\subsection{``Weak values'' are not unique, even in an AAV-type context}
We have seen that applying $(V \tnr I)e^{-iAP}$
to a starting product state $s \tnr m$ 
to obtain a state to be measured and postselected results in a ``weak value''
given by \re{eq600}, 
\beq
\lbl{eq630}
 \Re \frac{\lb f, VAs \rb}{\lb f, Vs \rb } =  
\Re \frac{\lb f, VAs \rb }{\lb f, s \rb }  
\q,   
\eeq
where we have finally used the assumption $Vs = s$, which will be shown
in Appendix 2 to guarantee ``weakness'' of the 
measurement.  This does not look like the AAV-type ``weak value'' of
equation \re{eq620}, namely
\beq
\lbl{eq640}
 \Re \frac{\lb f, As \rb}{\lb f, s \rb }
\q,
\eeq
but for logical completeness, we should check that they can
be numerically different, as well as different in appearance.  
The calculational argument given at the end of Section 5 is easily
adapted to show this, but for the reader's convenience and for variety,
we give here a more general argument which avoids calculation.  

Suppose the two  
``weak values'' given
by equations \re{eq630} and \re{eq640} are always equal.
Then 
\beq
\lbl{eq650}
\Re \frac{\lb f, VAs \rb }{\lb f, s \rb } = 
 \Re \frac{\lb f, As \rb}{\lb f, s \rb } 
\eeq 
for all states $s, f \in S$ with $\lb f, s \rb \neq 0$, all Hermitian
$A: S \rightarrow S$, and all unitary $V: S \rightarrow S$ satisfying 
$Vs = s$.  

Take $A$ to be any Hermitian operator with $As$ not a scalar multiple of
$s$, i.e., $s$ is not an eigenvector of $A$.  Let $u:= As - \lb s, As \rb s$
denote the component of $As$ orthogonal to $s$.  Then for any 
$w \in S$ with $|w| = |u|$ and $\lb w, s \rb = 0$, 
there exists a unitary $V$ on $S$ with 
$Vs = s $ and $Vu = w$, i.e., $VAs = w + \lb s, As \rb s$, and  
$$
\Re \frac{\lb f, VAs \rb}{\lb f, s \rb} = 
\Re \frac{\lb f, w \rb}{\lb f,  s \rb } + \lb s, As \rb \q. 
$$
By varying $w$, we can change 
$\Re (\lb f, VAs \rb / \lb f, s \rb $ 
except in the  special case in which $f$ is a multiple of $s$.
\section{Relation of the AAV approach to ours}
As mentioned earlier, our method of obtaining weakness of the measurement
differs from that of AAV.  This section notes that the difference is 
only cosmetic, and it also notes what we suspect may be an essential error
in the mathematics of AAV.

We obtain weakness of the meter measurement by  
replacing the state preparation
Hamiltonian $A \tnr P$ with $H(\epsilon) := \epsilon A \tnr P$.  This makes the
normalized (i.e., divided by $\epsilon$) 
expectation of the meter measurement conditional on postselection to $f$
(cf. \re{eq400}) equal to
\beq
\lbl{eq700}
\frac{\frac{1}{\epsilon}
\lb e^{-iH(\epsilon)} (s \tnr m), (P_f \tnr Q)
e^{-i H(\epsilon)} (s \tnr m) \rb
}
{\lb e^{-iH(\epsilon)} (s \tnr m), (P_f \tnr I) 
e^{-i H(\epsilon)} (s \tnr m) \rb
}.
\eeq
By contrast, AAV attempts%
\footnote{We say ``attempts'' not to cast doubt on the ``weakness'' of 
their procedure, but because they do not prove weakness, nor even discuss
it.  
}
to obtain weakness by using a fixed preparation Hamiltonian
$ A \tnr P$  (multiplied by an inessential constant) and instead replacing
our fixed meter state $m$ by an $\epsilon$-dependent meter state 
$m_{AAV}[\epsilon](\cdot)$ defined for $\epsilon > 0$ 
by  
\beq
\lbl{eq705}
m_{AAV}[\epsilon](q) := 
\left[
\frac{\exp (-\epsilon^2 q^2/ 2)
} 
{
\sqrt{2 \pi}/\epsilon
}
\right]^{1/2}
\q,
\eeq
which makes $m_{AAV}(\epsilon)^2$ a Gaussian centered at 0 with 
variance $1/\epsilon^2$.
Here we are translating AAV into our notation.  A reader consulting AAV
should remember that they interchange $P$ and $Q$ relative to our convention;
i.e., their ``meter'' is $P$, and their preparation 
Hamiltonian is $- A \tnr Q$ instead of our $A \tnr P$.

To see the relation of the two approaches, let $m$ be a fixed meter state
(not necessarily a Gaussian as in AAV) satisfying the conditions $1-4$
of Subsection 7.2, and for $\epsilon > 0$ define a new meter state
$m[\epsilon](\cdot) \in L^2(\bR)$ by 
\beq
\lbl{eq710}
m[\epsilon](q) := m(q\epsilon)\sqrt{\epsilon}
\q.
\eeq
The graph of $m[\epsilon](\cdot)$ is the graph of $m$ expanded by 
a factor of $1/\epsilon$ and then normalized to make its $L^2$ norm 
$||\, m[\epsilon]\, ||$ equal to 1.  For $m := m_{AAV}[1]$ as 
given by \re{eq705}
with $\epsilon := 1$, $m[\epsilon] $ coincides with $m_{AAV}(\epsilon)$ for 
arbitrary $\epsilon > 0$. 

We shall show that \re{eq700} can be rewritten as 
\begin{eqnarray}
\lbl{eq720}
\lefteqn{
\frac{\frac{1}{\epsilon}
\lb e^{-iH(\epsilon)} (s \tnr m), (P_f \tnr Q)
e^{-i H(\epsilon)} (s \tnr m) \rb}
{\lb e^{-iH(\epsilon)} (s \tnr m), (P_f \tnr I) 
e^{-i H(\epsilon)} (s \tnr m) \rb},
}\nonumber\\
&=&
\frac{
\lb e^{-iH(1)} (s \tnr m[\epsilon]), (P_f \tnr Q)
e^{-i H(1)} (s \tnr m[\epsilon] \rb
}
{ \lb e^{-iH(1)} (s \tnr m[\epsilon]), (P_f \tnr I) 
e^{-i H(1)} (s \tnr m[\epsilon]) \rb}.
\end{eqnarray}
In other words, the result of using the fixed Hamiltonian $H(1)$
with the $\epsilon$-dependent meter state $m[\epsilon]$ is the 
same as using our $\epsilon$-dependent Hamiltonian $H(\epsilon)$
with our fixed meter state $m$, and then normalizing by dividing by 
$\epsilon$.

We shall show that the numerator of the left side of \re{eq720}
equals the numerator of the right side, 
leaving the similar calculation of the denominators to the reader.
The notation will be that of the previous calculation of the AAV-type
weak value; in particular,
$s = \sum_i s_i$ is the decomposition of $s$ as a sum of orthogonal 
eigenvectors $s_i$ of $A$
with respective eigenvalues $\alpha_i$,
$A s_i = \alpha_i s_i $, and for $g \in L^2(\bR)$, 
$g_\beta (q) := g(q - \beta)$. 
Recalling that 
$$
e^{-iH(1)}(s\tnr g) = \sum_i s_i\tnr g_{\alpha_i} 
\q,
$$ 
we have, setting $q^\prime:= \epsilon q$, 
\begin{eqnarray}
\lbl{eq730}
\lefteqn{
\lb e^{-iH(1)}s \tnr m[\epsilon], (P_f \tnr Q) (s \tnr m[\epsilon]) \rb
} \nonumber\\
&=& 
\sum _{i,j} \lb s_i, P_f s_j \rb 
\int_{-\infty}^{\infty} \,  m[\epsilon]_{\alpha_i}(q)\, q \, 
m[\epsilon]_{\alpha_j}(q)\, dq
\nonumber\\
&=&
\sum _{i,j} \lb s_i, P_f s_j \rb 
\int_{-\infty}^{\infty} \,  m(\epsilon(q-\alpha_i))\, q \,  
m(\epsilon(q - \alpha_j) (\sqrt{\epsilon})^2  \, dq
\nonumber\\
&=&
\sum _{i,j} \lb s_i, P_f s_j \rb 
\int_{-\infty}^{\infty} \,  m(q^\prime- \epsilon \alpha_i))  (q^\prime/\epsilon) 
m(q^\prime - \epsilon\alpha_j) \epsilon \, dq^\prime/\epsilon
\nonumber \\
&=&
\frac{1}{\epsilon}
\sum_{i,j} \lb s_i, P_f s_j \rb \lb m_{\epsilon \alpha_i}, 
Q m_{\epsilon \alpha_j} \rb,
\end{eqnarray}
which is identical to the the numerator of \re{eq700} 
as calculated in \re{eq540}, \re{eq550}, and the following equations.

Since the AAV-setup seems essentially 
mathematically identical to ours, one  may wonder how the two
approaches end with different weak values, namely 
$\lb f, As \rb / \lb f, s \rb$ for AAV and the real part of that for us. 
Few readers will want to slog through the mostly routine but tedious 
mathematics of both to look for errors, so it may be helpful to point out
what we think might be an essential error in AAV.  

It is well known and routine to calculate  
that if $g = g(q)$ is a function in $L^2(\bR)$ and 
$\tilde{g} = \tilde{g}(p)$ denotes its Fourier transform, 
then multiplying $g(q)$
by $e^{i \alpha q}$, with $\alpha $ real, translates its Fourier transform
by $\alpha$.  That is, if $h(q) := e^{i\alpha q} g(q)$, then 
$\tilde{h}(p) = \tilde{g}(p - \alpha)$.  But this holds only for {\em real}
$\alpha$; for nonreal $\alpha$, $e^{i\alpha q}$ grows exponentially for 
large $|q|$ and $h$ cannot be expected to even have a Fourier transform,
since it cannot be expected to be in $L^2$.  But in passing from their 
fundamental equation (2) (via their uncontrolled approximation (3)) to 
their (5), AAV seems to assume the just-mentioned fact about Fourier transforms
for complex $\alpha$.  We suspect that this may be the origin of the difference
between their ``weak value'' and ours. 
\section{Another way to obtain non-standard ``weak values''}
We present another way to obtain ``weak values'' which differ from the usual
formulas $\Re (\lb f, As\rb /\lb f, s\rb)$ or 
$\lb f, As\rb /\lb f, s\rb.$  It is motivated by a comment on the Internet
newsgroup sci.physics.research which noted that the unitary operator $V$
appearing in the previous examples depends on the initial state $s$.   
But what if the initial state is not known?
The commentor (who posts under the pseudonym ``student'' and whom I thank)     
thought that it would be a rather peculiar type of measurement
that would not work for all states.

We agree that this does complicate the situation, but not fatally.  
First of all, it is frequently the case that the initial state {\em is}
given by the physical situation; for example this is the case for the 
investigations of ``Hardy's paradox'' in \cite{RevHardy} and \cite{Yokota}.
The nonuniqueness of weak values for such situations is a counterexample 
to claims that weak values must be given by the above ``usual'' formulas.

Second, in all formulations of weak measurements known to us, the initial
state $s$ has to be assumed to be ``reproducible'', i.e., the experimenter
must have an apparatus which will produce any number of copies of $s$.
This is because for very weak coupling between the system of interest $S$ 
and the meter system, a large number of meter measurements may be required
to obtain a reliable average.  And in finite dimensins, 
a reproducible state  {\em is} effectively
known because its components with respect to any given basis 
can be determined 
to arbitary accuracy by quantum tomography (\cite{N/C} pp. 389 ff).  
If we want to make a weak
measurement involving a starting state $s \in S$, we can first calculate its components and then
construct the desired operator $V$ (to arbitrary accuracy) using quantum
gates (\cite{N/C}, Chapter 4).

However, the comment did cause us to look for other ways 
which did not depend explicitly on the initial state
to obtain non-traditional  weak values.  One such way was
already known from our study of the Yokota, et al. paper 
\cite{Yokota} (as described in the next section), but this seemed 
undesirably complicated for readers without 
detailed knowledge of that  paper.  

The result of the search seemed surprising.  We shall present below
a way to obtain weak values 
such that the preparation Hamiltonian (the analog of $H(\epsilon) := 
\epsilon A \tnr P$ of \re{eq200}) 
does not depend on the initial state $s\in S$.  The preparation Hamiltonian
used is very similar to the \re{eq200} 
used by von Neumann and AAV except that it
requires only a finite dimensional meter space (which can be as small
as dimension 2), instead of the infinite
dimensional meter space $L^2(\bR)$ used by them.  
Since we continue to assume that the system $S$ of interest is finite 
dimensional, the finite dimensionality of the meter space means that the 
calculations will be purely algebraic, and rigorous.  
The algebra will be identical for an infinite dimensional meter space,  
though further considerations would be 
necessary to make the calculation rigorous.  

The general method will be identical to that used in Subsection 7.2
to calculate the AAV-type ``weak value''.  The difference is that instead
of the preparation Hamiltonian $H(\epsilon) := \epsilon A \tnr Q$ used 
there,  
we shall use a preparation Hamiltonian of the form
\beq
\lbl{eq800}
H(\epsilon) := \epsilon A \tnr G
\eeq
with $G: M \rightarrow M$ Hermitian.    The Hermitian operator on $M$
whose value gives the meter reading will be called $B: M \rightarrow M$
instead of the $Q$ used in Subsection 7.2.  
Thus $G$ plays the role of $P$ and $B$ the role of $Q$ in Subsection 7.2.

A starting state $s \tnr m$ is assumed given, with $|s| = 1 = |m|$ and 
\beq
\lbl{eq810}
\lb m , Bm \rb = 0
\eeq
as before.
The state to be 
meter-measured will be obtained as before by applying $e^{-i H(\epsilon)}$
to $s \tnr m$:
\beq
\lbl{eq820}
e^{-i\epsilon (A \tnr G)}(s \tnr m) = s \tnr m - i \epsilon As \tnr Gm + O(\epsilon^2)
\q. 
\eeq
Then the normalized (by division by $\epsilon$) average value of the
meter measurement is 
\begin{eqnarray}
\lbl{eq830}
\lefteqn{
\frac{\lb e^{-i \epsilon (A \tnr G)}(s \tnr m), 
(I \tnr B) e^{-i \epsilon (A \tnr G)}(s \tnr m) \rb}{\epsilon}
}
&& \nonumber \\
&=&   
\frac{\lb s, s \rb  \lb m , Bm \rb
- i \epsilon ( \lb s , As \rb \lb m, BGm \rb - \lb As, s \rb \lb Gm, Bm \rb)
+ O(\epsilon^2)}
{\epsilon} 
\nonumber \\
&=& 
-i  \lb s, As \rb \lb m, BG - GB m\rb + O(\epsilon)
\q,
\end{eqnarray} 
and
\begin{eqnarray}
\lbl{eq840}
\lefteqn{
\lim_{\epsilon \goesto 0} 
\frac{\lb e^{-i \epsilon (A \tnr G)}(s \tnr m), 
(I \tnr B)\lb e^{-i \epsilon (A \tnr G)}(s \tnr m) \rb}
{\epsilon} 
}
&& \nonumber\\
&=& -i \lb m, (BG - GB) m \rb \lb s, As \rb
.
\end{eqnarray}
Thus if we choose $m, B$, and $G$ such that 
\beq
\lbl{eq845}
1 = -i \lb m,  (BG - GB)m \rb = 2 \Im \lb m, BGm \rb
\q,
\eeq 
measuring $B$ in the meter space and normalizing 
by dividing by $\epsilon$ (for small $\epsilon$) 
constitutes a weak measurement protocol in the sense of Definition 1.
(We omit the proof that the measurement negligibly affects the state of $S$
for small $\epsilon$, which is identical to that given in Subsection 5.3.)

The average value of the normalized meter reading conditional on successful 
postselection to $f \in S$ is
\begin{eqnarray}
\lbl{eq850}
\lefteqn{
\frac{1}{\epsilon} 
\frac{\lb e^{-i \epsilon (A \tnr G)}(s \tnr m), 
(P_f \tnr B)\lb e^{-i \epsilon (A \tnr G)}(s \tnr m) \rb
}
{\lb e^{-i \epsilon (A \tnr G)}(s \tnr m), 
(P_f \tnr I)\lb e^{-i \epsilon (A \tnr G)}(s \tnr m) \rb
}
}
&&\nonumber\\
&=& 
\frac{-i ( \lb s, P_f As \rb \lb m , BG m \rb - \lb As, P_f s \rb 
\lb Gm, Bm \rb ) + O(\epsilon)}
{\lb s , P_f s \rb + O(\epsilon)} 
 \nonumber \\
&=& 
\frac{ \lb f, As \rb \lb s, f \rb \lb  m, BG m \rb -
\lb As, f \rb \lb f , s \rb \lb m, GB m \rb 
}
{\lb f, s \rb \lb s , f \rb} + O(\epsilon)
\nonumber\\
&=&
-i \left[ \frac{\lb f, As \rb \lb m, BGm \rb}{\lb f, s \rb}
- \frac{ \lb As, f \rb \lb BGm, m \rb}{\lb s, f \rb} \right] 
+ O(\epsilon)
\nonumber\\
&=& 2 \Im \frac{ \lb f, As \rb \lb m, BGm \rb}{\lb f, s \rb} 
+ O(\epsilon) 
\end{eqnarray}    
when $\lb f, s \rb \neq 0$  and undefined if $\lb f, s \rb = 0$.  The 
following calculations assume $\lb f , s \rb \neq 0$.
Write 
\beq
\lbl{eq 860}
\lb m, BGm \rb = \rho + i \kappa  \q \mbox{with $\rho, \kappa$ real.}
\eeq 
From \re{eq845} $\kappa = 1/2$.
Then the limit as $\epsilon \goesto 0$ of \re{eq850} becomes
\begin{eqnarray}
\lbl{eq860}
\lefteqn{
\lim_{\epsilon \goesto 0}
\frac{1}{\epsilon} 
\frac{\lb e^{-i \epsilon (A \tnr G)}(s \tnr m), 
(P_f \tnr B)\lb e^{-i \epsilon (A \tnr G)}(s \tnr m) \rb
}
{\lb e^{-i \epsilon (A \tnr G)}(s \tnr m), 
(P_f \tnr I)\lb e^{-i \epsilon (A \tnr G)}(s \tnr m) \rb
}
} 
\mbox{\ \ \ \ \ \ \ \ \ \ \ \ \ \ \ \ \ \ \ \ \ \ \ \ \ \ \ \ \ \ \ \ \ \ \
\ \ \ \ \ } 
&&\nonumber\\
&=& 2\kappa \Re \frac{\lb f, As \rb}{\lb f, s \rb} + 2\rho \Im
\frac{\lb f, As \rb}{f , s \rb} \nonumber\\
&=&  
 \Re \frac{\lb f, As \rb}{\lb f, s \rb} + 2 \rho \Im
\frac{\lb f, As \rb}{f , s \rb} \nonumber\\
.
\end{eqnarray}
The first term in \re{eq860} is the ``usual'' weak value
$\Re( \lb f, As \rb / \lb f, s \rb)$, but when $\Im \lb f, As \rb
\lb f, s \rb \neq 0,$ we shall show that the second term can be chosen
arbitrarily by adjusting the value of $\rho = \Re \lb m, BGm \rb$.    

Recall that $m, B$, and $G$ are arbitrary subject to 
$\Im \lb m, BG m \rb = 1/2$ and $\lb m, B m \rb = 0$. 
To see that any number can be obtained for $\Re \lb m, BGm \rb$ under 
these conditions, take the meter space $M$ to be two-dimensional
with orthonormal basis $m, m^\perp$, and define $G$ and $B$ by the
following matrices with respect
to this basis:
\beq
\lbl{eq870}
G := 
\left[
\begin{array}{ll}
0 & 1  \\
1  & 0
\end{array}
\right]
\q \mbox{and}\q 
B := 
\left[
\begin{array}{ll}
0 & \rho + i/2 \\
\rho - i/2 & 0
\end{array}
\right]
\mbox{with $\rho$ real.} 
\eeq
Then $BG$ has the following form, where entries denoted ``$*$'' have not
been calculated because they are irrelevant to calculation of 
$\lb m , BGm \rb$ (which is the upper left entry of BG):
\beq
\lbl{eq880}
BG = 
\left[ 
\begin{array}{rr}
\rho + i/2 & * \\
** & * \\
\end{array} 
\right]
= 
\left[ 
\begin{array}{rr}
\lb m, BGm \rb & * \\
** & * 
\end{array}
\right].
\eeq
When $\Im (\lb f, As \rb / \lb f, s \rb) \neq 0$, 
by varying $\rho$, one can obtain any number whatever as a ``weak value''
for $A$.

The above calculations are rigorous for finite-dimensional $S$ and $M$, and 
still algebraically correct in infinite dimensions.  In the physics literature,
such algebraic calculations are typically accepted as ``proofs''.
If we relax mathematical rigor to this extent, we obtain from the above
a very simple ``proof'' of the ``usual'' weak value $\Re (\lb f, As \rb / 
\lb f, s \rb)$ by taking 
\beq
\lbl{eq890}
M := L^2(\bR), \q  G := P,\q  B := Q , \q m(q) := 
\left[\frac{1}{\sqrt{2\pi}}
e^{- q^2/2}\right]^{1/2} ,
\eeq
where $P$ and $Q$ are the usual momentum and position operators, respectively, 
defined in Subsection 6.2.   

This ``proof'' is deeply flawed because the starting
equation \re{eq820},
$$
e^{-i\epsilon (A \tnr G)}(s \tnr m) 
= s \tnr m - i \epsilon As \tnr Gm + O(\epsilon^2)
,
$$
would probably be difficult to justify for our unbounded $G := P$. 
Some ``proofs'' of the ``usual'' weak value formula in the literature
rely on uncontrolled approximations like this.  In honesty, they should be
called something like ``algebraic motivations'' instead of proofs. 

If we {\em are} willing to accept uncritically such uncontrolled approximations,
we can obtain arbitrary weak values in an AAV-type framework by 
taking $ B$ and $m$ as in equation \re{eq890} and 
\beq
\lbl{eq892}
G : P + \delta Q \q \mbox{ with $\delta \neq 0$ real.} 
\eeq
This results in 
\beq
\lbl{eq894}
\lb m , BG m \rb = \delta  + \frac{i}{2} \q,
\eeq
so that when $\Im ( \lb f, As \rb / \lb f , s \rb) \neq 0$, any ``weak value''
whatever can be obtained using the preparation Hamiltonian 
$H(\epsilon) := \epsilon A \tnr (P + \delta Q)$.   

Since $\delta$ can be arbitrarily small, anyone who claims that the ``usual''
weak value formula gives the only experimentally possible result 
should be obligated to explain how the von Neumann/AAV Hamiltonian
$H(\epsilon) := \epsilon A \tnr P$ can be guaranteed in any experimental
situation, and assuming that, how the Hamiltonian $A \tnr P$ can be 
experimentally distinguished from $A \tnr (P + \delta Q)$ for arbitrarily 
small $\delta$. On the level of uncontrolled approximations, this will 
probably be impossible.  

It is almost immediate that  
\beq
\lbl{eqS820}
e^{-i Q^2 \delta/2} P e^{i Q^2 \delta/2} = P + \delta Q 
\eeq
because for any $g \in L^2 (\bR)$, $(e^{-i Q^2 \delta/2}g)(q)
= e^{-i q^2 \delta/2}g(q)$ and $P := -i d/dq$.  (We say ``almost'' because
a rigorous verification would
require careful specification of the domain of $P$, which
we have not discussed.)  In other words, the new Hamiltonian 
$A \tnr (P + \delta Q)$
is formally unitarily equivalent to the AAV Hamiltonian $A \tnr P$.

Thus the above setup with the new Hamiltonian can be unitarily transformed into
one with the AAV Hamiltonian.  The transformation will take the AAV meter 
state $m$ of equation \re{eq890} into the new meter state 
\beq \lbl{eqS826} 
q \mapsto e^{-iq^2 \delta /2} m(q)
\q, 
\eeq
whose absolute square 
defines the same Gaussian probability distribution on position space as $m$.  
This suggests that it might be possible to rework the rigorous calculation
of the traditional weak value $\Re (\lb f, As \rb / \lb f, s \rb)$ in 
Subsection 7.2 into
a rigorous calculation which gives arbitrary weak values 
in a slightly different setup using the AAV Hamiltonian $A \tnr P$, but  
the slightly different meter state \re{eqS826}.  The next section will 
carry this out.  
\section{Another way to rigorously obtain non-standard ``weak values''
in a framework similar to AAV}
The rigorous calculation of the traditional weak value 
$\Re \lb f, As \rb / \lb f, s \rb)$  of Subsection 7.2 
assumed that the meter state $m \in L^2(\bR) $ is a  real-valued function.  
After that proof was typeset, we have noticed that without that assumption, 
the proof of that section is easily adapted 
to {\em rigorously} obtain
non-traditional (in fact, arbitrary) 
weak values in a setting very similar to AAV.
More precisely, it assumes the AAV preparation Hamiltonian with a non-real 
meter state which is the AAV meter state (the square root of a Gaussian) 
multiplied by a complex function of modulus 1, so the new meter state 
defines the same Gaussian probability distribution in position space as
the AAV meter state.  This section will outline the necessary modifications
to the argument of subsection 7.2.  We continue to assume that the meter 
state $m$ satisfies the normalization and growth conditions 2, 3, and 4 of
Section 7.2, but we no longer assume condition 1 that it be real.

The  new proof is identical to the old through equation \re{eq540}, written
here with a new number, and with $V := I$: 
\begin{eqnarray}
\lbl{eqN540}
\lefteqn{\lb  e^{-iH(\epsilon)}(s \tnr m), (T \tnr Q)
e^{-iH(\epsilon)} 
	(s \tnr m) \rb} \nonumber\\
&=& 
\lb \sum_i s_i \tnr m_{\epsilon \alpha_i} ,
 (T\tnr Q) \sum_j s_j \tnr m_{\epsilon \alpha_j} 
\rb \nonumber\\
&=&
\sum_{i,j}  
\lb s_i , Ts_j \rb \lb m_{\epsilon \alpha_i} , (Qm)_{\epsilon \alpha_j} \rb 
\q.
\end{eqnarray}
The inner product involving $m$ is 
\begin{eqnarray}
\lbl{eqN550}
\lb m_{\epsilon \alpha_i} , Qm_{\epsilon \alpha_j} \rb 
&=& \int_{q = -\infty}^\infty \, 
m^*(q - \epsilon \alpha_i)\, q \, m(q - \epsilon \alpha_j) \, dq
\nonumber\\
&=& 
\int \, m^* (q ) (q + \epsilon \alpha_i) 
m(q - \epsilon (\alpha_j - \alpha_i) )\, dq
\nonumber\\ 
&=& 
\int \, m^*(q)\epsilon \alpha_i m(q- \epsilon (\alpha_j - \alpha_i)) \, dq
\nonumber\\
&& +    
\int \, m^*(q)\, q \, [m(q) - m^\prime (q) \epsilon(\alpha_j - \alpha_i)] 
\, dq  +  O(\epsilon^2)
.
\end{eqnarray}
The first integral in \re{eqN550} is the same as before, namely
$\epsilon (\alpha_i + o(1))$, but integration by parts only determines
the real part of the second integral, which is the same as before, 
namely $\epsilon (\alpha_i - \alpha_j)/2.$  

We shall not evaluate the imaginary part of the second integral in general,
but only for a specific meter function $m$ which we assume to be of the form
\beq
\lbl{eqN560}
m(q) = e^{-i q^2 \delta /2} m_0 (q) \q, 
\eeq 
where $\delta $ is a real constant and $m_0$ a real meter function satisfying
the same conditions as assumed in Subsection 7.2%
\footnote{ 
Those are that $m_0$ be normalized, i.e., 
$||m_0|| = 1$, and satisfy the previous condition 2 
that $\lb m_0, Q m_0 \rb$, and the growth conditions 3 and 4.
}
 together with the 
additional normalization condition
\beq
\lbl{eqN565}
\lb m_0, Q^2 m_0 \rb = 1 \q.
\eeq
For example, we could take $m_0$ as the AAV meter function $m_{AAV}[1]$, 
defined in equation \re{eq705} as the square root of
a Gaussian with mean 0 and variance 1.

Then, since 
$$
m^\prime (q) = e^{-i q^2 \delta/2}[m^\prime_0 (q) - i \delta q m_0(q)]
\q,
$$ 
\begin{eqnarray}
\lbl{eqN570}
\lb m_{\epsilon \alpha_i} , Qm_{\epsilon \alpha_j} \rb 
&=& \int_{q = -\infty}^\infty \, 
m^*(q - \epsilon \alpha_i)\, q \, m(q - \epsilon \alpha_j) \, dq
\nonumber\\
&=& \epsilon (\alpha_i  + o(1)) + 
+ \epsilon \frac{\alpha_j - \alpha_i}{2} + O(\epsilon^2)
\nonumber\\
&&
+ \epsilon  (\alpha_j - \alpha_i) 
\int_{-\infty}^{\infty}  \, m_0 (q) q^2 m_0 (q) \delta i \, dq
\nonumber \\
&=&  
 \epsilon \frac{\alpha_j + \alpha_i}{2} + 
\epsilon (\alpha_j - \alpha_i)\delta i + \epsilon o(1) 
.
\end{eqnarray} 
This is the same result as for a real meter function plus the imaginary
part $(\alpha_j - \alpha_i ) \delta i$.  
(We assume that the reader will take in stride the use of the same symbol
$i$ for the imaginary unit and an index as in $\alpha_i$.  Since the previous
version only used real quantities, no ambiguity arose there, and changing
notation for the index at this point probably risks more confusion
than retaining it.) 

The main step to complete the calculation is to  
evaluate the normalization of 
\re{eqN540} in the limit $\epsilon \goesto 0$.  There are two cases
to consider, $T := I$ to demonstrate that the normalized meter 
expectation does equal $\lb s, As \rb $
and $T := P_f$ to calculate the weak value.  Here we take $T := P_f$,
leaving the case $ T := I$ to the reader. We have
\begin{eqnarray}
\lbl{eqN580}
\lefteqn{
\lim_{\epsilon \goesto 0} 
\frac{1}{\epsilon}
\lb ( e^{-iH(\epsilon)}(s \tnr m), (P_f \tnr Q)
e^{-iH(\epsilon)} 
	(s \tnr m) \rb} \nonumber\\
&=&
\sum_{i,j}  
\lb s_i , P_f s_j \rb \lb m_{\epsilon \alpha_i} , (Qm)_{\epsilon \alpha_j} \rb 
\nonumber \\
&=&
\sum_{i,j} \lb s_i, P_f s_j \rb \left[ \frac{\alpha_i + \alpha_j}{2}
+ (\alpha_j - \alpha_i) \delta i \right]
\nonumber \\
&=& \Re \lb As, P_f s \rb - 2\delta \Im \lb As, P_f s \rb 
\q.
\end{eqnarray}
To understand the passage to the last line, first recall that 
the $s_i$ are  eigenvectors of $A$ with $As_i = \alpha_i s_i$,
and $s = \sum_i s_i$.
We shall concentrate on evaluating
$$
\sum_{i,j} \lb s_i , P_f s_j \rb (\alpha_j - \alpha_i)\q,  
$$
leaving the other term, which is similar 
and  was effectively done in the original proof,
to the reader. We have
\begin{eqnarray*}
\lbl{eqN590}
\sum_{i,j} \lb s_i , P_f s_j \rb (\alpha_j - \alpha_i)
&=& \sum_i \sum_j \lb s_i , P_f s_j \rb (\alpha_j - \alpha_i)
\nonumber\\
&=& \sum_i [\lb s_i , P_f As \rb -  \alpha_i \lb s_i , P_f s \rb]
\nonumber \\
&=& \lb s, P_f As \rb -  \lb As , P_fs \rb 
\nonumber \\
&=& \lb P_f s , As \rb - \lb As, P_f s \rb
\nonumber \\
&=& 2i \Im \lb P_f s , As \rb
\q.
\end{eqnarray*}
The desired weak value is 
\beq 
\lbl{eqN605}
\lim_{\epsilon \goesto 0} \frac{1}{\epsilon}
\frac{\lb 
e^{-iH(\epsilon)}(s \tnr m), (P_{f} \tnr Q)e^{-iH(\epsilon)} 
	(s \tnr m) \rb
	}
	{ \lb  e^{-iH(\epsilon)}(s \tnr m), (P_f \tnr I)
e^{-iH(\epsilon)}(s \tnr m) \rb 
	}
\eeq
The limit as $\epsilon \goesto 0$ of the 
denominator of the second fraction was effectively computed in 
\re{eq530} (which did not require the reality of the meter function) 
as $\lb f, s \rb \lb s, f \rb$.
Combining this with \re{eqN580} gives the weak value as 
\begin{eqnarray}
\lbl{eqN610}
\lefteqn{
\lim_{\epsilon \goesto 0}
\frac{\lb 
e^{-iH(\epsilon)}(s \tnr m), (P_{f} \tnr Q)e^{-iH(\epsilon)} 
	(s \tnr m) \rb
	}
	{\epsilon \lb  e^{-iH(\epsilon)}(s \tnr m), (P_f \tnr I)
e^{-iH(\epsilon)}(s \tnr m) \rb 
	}
}\nonumber \\
&=&  
\frac{\Re\lb As, P_f s \rb - 2 \delta \Im \lb As, P_fs \rb
}
{
\lb f, s \rb \lb s, f \rb} 
\nonumber \\
&=& \Re \frac{\lb As, f \rb}{\lb s, f \rb} 
- 2 \delta \Im \frac{\lb As, f \rb}{\lb s, f \rb} 
\nonumber \\
&=& \Re \frac{\lb f, As \rb}{\lb f , s \rb}
+ 2 \delta \Im \frac{ \lb f, As \rb}{\lb f, s \rb}
\q.
\end{eqnarray}

This shows {\em rigorously} that when 
$\Im (\lb f, A s \rb / \lb f , s \rb) \neq 0$,
any weak value whatever may be obtained by some postselected weak measurement
protocol using the AAV preparation Hamiltonian and a meter state different
from that used by AAV, but which defines the same Gaussian probability 
distribution on position space. 

Therefore, any argument that the ``usual'' weak value 
$ \Re (\lb f, As \rb / \lb f, s \rb)$ is experimentally inevitable will
probably have to argue for the experimental universality of the
precise AAV setup 
(meter observable, meter state, and  preparation Hamiltonian).  Since 
there seems no argument in the literature that this precise setup can be
realized in {\em any} experimental situation,%
\footnote{Since there have been experiments which do obtain the ``usual''
weak value, that might be considered suggestive that the AAV framework 
can be experimentally realized.  However, 
we have seen that other frameworks can produce
the same weak value.  We have never seen an argument in the experimental
literature that the experimental situation actually does implement 
the AAV setup. 
} 
much less {\em all}, such an argument would probably have to break new
ground.  
\section{Rambling afterword with conclusions}

After spending months learning about weak values in
the hope that it might lead to some fundamental new understanding of 
quantum mechanics, I was disappointed that this hope did not materialize.
This work was written mainly to save others in a similar situation from 
similarly wasting their time.

I do not recall the exact route I followed through the ``weak values'' 
literature, but it took me through all of it that I could find. 
I think the first paper I read might have been a very interesting short
paper of Yokota, {et al.}, which describes an experimental measurement of
weak values to confirm a quantum mechanical prediction concerning
``Hardy's Paradox''.%
\footnote{``Hardy's Paradox'' is what it is commonly called in the literature,
but Hardy \cite{hardy} did not present it as a paradox, and I think it
questionable to call it that.
}
Then I probably looked at the mostly clear and interesting analysis of 
``Hardy's Paradox'' in \cite{RevHardy}, followed by carefully reading 
AAV \cite{AAV}, but it could have been in the reverse order.  I have made
no attempt to reference all the papers in my thick folder on weak values.

I was struck by the fact that all the literature which I saw gave 
the impression that one of the two versions of the ``standard'' formulas 
for weak values, equation \re{eq242}, $\Re (\lb f, As \rb/\lb f, s \rb$,
 or \re{eq245}, $\lb f, As \rb / \lb f, s \rb$ was a universal formula
to be expected in all experimental situations.  But AAV's motivation of 
\re{eq245} did not seem fundamental because it relied on assumption of the
particular Hamiltonian \re{eq200}, without any discussion of how to guarantee
that Hamiltonian in {\em any} experimental situation, much less in {\em all}
experimental situations.  

Yokota, {\em et al.} \cite{Yokota} obtained the 
standard weak values formula \re{eq242} using a Hamiltonian completely 
different from that of AAV or other motivations for \re{eq242} that I have
seen. For example, AAV and Lundeen/Steinberg's \cite{LS} both use
 \re{eq200}, which employs an infinite dimensional meter space, while 
Yokota, {\em et al.} use a four-dimensional meter space.  It seemed 
remarkable that these disparate approaches yielded the same formula.  

At first, it seemed inconceivable that they could unless they were both
reflecting aspects of some fundamental, as yet undiscovered physical 
reality or mathematical fact.  I set out to find an explanation for the 
fact that AAV and  Yokota, {\em et al.}, had obtained similar ``weak values''
with such utterly different methods and physical setups.  

The effort led to disappointment.  I did not discover any new physical or
mathematical principles.

  On careful reading of Yokota, 
{\em et al.}'s \cite{Yokota} analysis, I noticed some arbitrary elements.
Their method for obtaining weak values was not unreasonable, but it seemed
idiosyncratic, and  I would have done it differently.  I worked out 
the consequences of my method and did indeed obtain a different ``weak value''.
My method had no claim to being superior to theirs, but the exercise 
convinced me that weak values are not unique.

I did not present my method for obtaining weak values in the setup of
Yokota, {\em et al.} \cite{Yokota} in this paper because it is more complicated than
the nonuniqueness examples presented and would require extensive analysis 
of \cite{Yokota}.  The examples presented in this paper may seem artificial,
but they have the merit of being simple.

The fact that ``weak values'' different from those of  
Yokota, {\em et al.}, \cite{Yokota} can reasonably be obtained for their 
setup does not 
imply that there is anything wrong with their paper.  Their weak values 
and mine are both obtained using quantum mechanics so standard that it is
unlikely to be questioned.
So long as their experimental procedures accurately reflect the mathematical
methods used to obtain their weak values, their experiment can be viewed as 
a confirmation of standard quantum mechanics.  Their modestly written paper
does not claim anything further.  In particular, they (wisely) do not    
claim to have resolved ``Hardy's Paradox''.

The ``Reader's Guide'' section remarked that I suspect that the introduction
to weak values and the finite-dimensional examples of Section 5 may turn out
to be all that most readers care to learn about weak values.  That is because
the ideas seem so unremarkable when presented in that context and because
I think that the literature which makes them seem remarkable rests on an 
implicit logical fallacy exposed in that section.  But I have no illusions
that all readers will share these views.

Readers who do think it worth their while to read further will find what I hope  
is a mathematically rigorous derivation of the standard weak value formula.
To a non-mathematician it may look complicated, but that is mainly because
everything is precisely defined and much more fully written out than is usual
in the physics literature.  The fact that it applies to almost any 
real ``meter'' state, not just a Gaussian meter state as in the existing
literature, may be viewed as an advance by those who take seriously the 
universality of the Hamiltonian \re{eq200} which AAV and most of the existing
literature implicitly assume. 
\section{Closing remarks on notation}
The notation chosen may irritate some readers.  It is intended to stay 
as close as possible to notation common in the physics literature without
sacrificing mathematical accuracy (as physics notation often does).  It
necessarily involves some compromises.

Physicists who have never seen a proper introduction to tensor products may
find notation like $s \tnr m$ strange and excessive.  Its merit is that it
is precise, unlike physics-type notation like $|s\rb |m \rb$ or 
$|s\rb_S |m\rb_M$.  (Either would probably suggest to many readers
that $s$ is a label for 
an eigenvalue of some operator, which it is not.  And if it is already 
understood,
that $s$ is merely a label for a vector state in $S$, 
what additional information would be given by the brackets and subscript
of $|s\rb_S$?) 

The use of ``projector'' in place of the 
``projection'' nearly universal in the mathematics literature may grate
on some mathematicians.  Who does this dude think he is to give him 
the right to substitute
his own notation for what is standard?%
\footnote{A referee for my book on relativistic electrodynamics 
made a similar impassioned objection to my choice of
``$\perp$'' for the Hodge dual operation in place of the ``$*$'' which
most mathematical literature uses, and is usually called the ``Hodge $*$
(or star) operation''.   I had chosen ``$\perp$'' because 
it seemed more common in the physics literature after it was popularized
by its use in the monumental treatise {\em Gravitation} of Misner, Thorne,
and Wheeler (MTW) and because it more accurately reflects the geometric
content of the operation.  
I imagine that the referee was a pure mathematician  
who might never have seen MTW. 
}

I chose ``projector'' mainly 
because that is the term most common in the physics literature. It also 
seems to me the more grammatically appropriate alternative.  
If a ``bettor'' is one who bets, 
shouldn't a ``projector'' be  one who projects?
If a ``bettor'' produces a bet, it seems reasonable that a ``projector'' 
should produce a ``projection''.  The projector $P_u$ produces 
the projection $[u]$ ( the subspace spanned by $u$).
Mathematicians who find this unconvincing and are still offended by 
``projector'' may be mollified by the assurance that 
I intend to revert to ``projection'' instead of ``projector''
in my next purely mathematical paper! 

Finally, we have bowed to the tradition of the physics literature
by describing the time evolution
of a quantum system initially in state $\psi$ by 
$t \rightarrow e^{-iHt} \psi$, where $H$ is the Hamiltonian, instead of the 
simpler and more natural $t \mapsto e^{iHt}\psi$.  
The complication of the minus sign,
though slight in any given instance, 
is a constant annoyance to mathematicians.  In the complex
number system, $-1$ has two square roots, and ``$i$'' is an arbitrary label
for one of them.  The two square roots have identical algebraic properties,
so either of them could be called ``$i$''.  The convention 
$t \mapsto e^{-iHt}\psi$  suggests that ``$-i$'' is in some fundamental way
different from ``$i$'', and that it would be wrong to write the time evolution
as $t \mapsto e^{iHt}\psi$. 
\section{Appendix 1:  The partial trace}

The definition of trace class operators on infinite dimensional Hilbert spaces
is surprisingly subtle, as is the definition of partial trace.  
If one does not feel the need to worry about convergence and orders of summation, 
the definitions can be lifted verbatim from the corresponding definitions
for finite matrices.  However, since we are trying to perform rigorously
calculations which are done non-rigorously in the physics literature
and have led to serious errors, that approach would not fulfill our needs.

On the other hand, we do not want to devote pages to carefully developing 
the properties of traces.  We choose a middle approach stating the
properties of traces which we shall use, either giving references 
or indicating how they can be proved.  Good references for
partial traces seem particularly hard to find.  I would be grateful for
any that readers might furnish.  

Besides summarizing properties
of these traces which we shall need, this appendix  
performs some calculations which will be used in the main text.    
Our primary reference for the definition
of trace class operator is A.\ Knapp's book, {\em Advanced Real Analysis} 
\cite{knapp}.  It efficiently 
develops the properties of trace class operators on infinite dimensional
Hilbert spaces, but does not discuss the partial trace.

Below the term ``operator'' will always mean ``bounded operator on
a Hilbert space. 
A bounded operator $T$ is said to be
of {\em trace class} if for all orthonormal sequences 
$\{u_i\}^\infty_{i=1}$ and  
$\{v_i\}^\infty_{i=1}$, 
\beq
\lbl{eqA10}
\sum_{i=1}^\infty |\lb u_i, Tve_i \rb | < \infty \q. 
\eeq
Its {\em trace}, denoted $\tr T$, is defined as
\beq \lbl{eqA20}
\tr T :=  
\sum_{i=1}^\infty |\lb u_i, T u_i \rb | < \infty \q. 
\eeq
This sum can be shown to be independent of the orthonormal basis $\{u_i\}$.  
In finite dimensions, this is a routine algebraic calculation, but an 
efficient proof for infinite dimensions requires careful organization.

To define partial traces, it will be helpful to identify operators
with sesquilinear forms: an operator $T: H \rightarrow H$ is identified
with the sesquilinear form $Q_T(\, \cdot \, , \, \cdot \,)$ defined for 
all $u, v \in H$ by
\beq
\lbl{eqA30}
Q_T (u,v) := \lb u, Tv \rb \q.
\eeq
When naming this form would only be a distraction, we shall refer to it
as the form
\beq
\lbl{eqA40}
u, v \mapsto \lb u, Tv \rb \q.
\eeq  
It is well known (and easy to prove) that this provides a one-to-one 
correspondence between operators and bounded sesquilinear forms. 
This identification of operators with forms will allow us to read off 
properties of the partial trace from corresponding properties of the 
ordinary trace.

Let $S$ and $M$ be Hilbert spaces, $S \tnr M$ their tensor product,
 and $L : S \tnr M\rightarrow S \tnr M$ a trace class operator on this tensor
product. For given vectors $s, s^\prime \in S$, consider the form on $M$
defined for all $m, m^\prime \in M$ by
\beq
\lbl{eqA50}  
m, m^\prime \mapsto \lb s \tnr m , L(s^\prime \tnr m^\prime ) \rb 
\q.
\eeq 
For fixed $s, s^\prime$, this form corresponds to an operator on
$M$ which is easily seen to be trace class (because $L$ is trace class
on $S \tnr M$).  Its trace is the definition of  
the {\em partial trace with respect to $M$}
of $L$, denoted $\trM L$, as a sesquilinear form.  More concretely,
for any orthonormal basis $\{f_i\}$ for $M$, 
\beq
\lbl{eqA60}
\lb s , (\trM L) s^\prime \rb = \sum_{\alpha}  \lb s \tnr f_{\alpha}, 
L(s^\prime \tnr f_{\alpha}) \rb
\q. 
\eeq 
(We sometimes distinguish dummy indices of summation by Greek.)
From the assumption that $L$ is trace class, it follows easily that
the sum in the definition \re{eqA60} converges absolutely, and also that
$\trM L$ is trace class and $\tr (\trM L) = \tr L$.%
\footnote{
To see this, note that for normalized $s$ and $s^\prime$, 
the sequences $\{s\tnr f_i\}$ and $\{s^\prime  \tnr f_j\}$ 
are orthonormal, 
so the definition of ``$L$ is trace class on $S \tnr M$''
applies directly to assure the absolute convergence of the sum in 
\re{eqA60}.    
The other facts follow from similar observations.
} 

The definition \re{eqA60} can also be viewed more concretely 
as defining $\trM L$ as a matrix.  
If $\{e_k\}$
is an orthonormal basis for $S$, and if the matrix of $L$ with respect to 
the orthonormal basis $\{e_j \tnr f_k \}$ for $S \tnr M$ is $(L_{il,jk })$,
then the matrix for $\trM L$ is $(\sum_{\alpha} L_{i\alpha,j\alpha})$.  
An advantage of
giving the definition in terms of forms instead of matrices is that it 
avoids the nuisance of checking that it is basis-independent.   

Let $\{f_i\}$ be an orthonormal basis for $M$.  Then any unit vector 
$u \in S \tnr M$ can be written uniquely as 
\beq
\lbl{eqA70}
u = \sum_{\alpha} s_{\alpha} \tnr f_{\alpha} \q \mbox{with} 
\sum_{\alpha} |s_{\alpha}|^2 = 1 \q.
\eeq
Recall that $P_u$ (the projector onto $u$) 
represents the mixed state (often called a 
density matrix on $S \tnr M$) corresponding to the vector $u$. 
We shall derive a revealing formula for the mixed state $\trM P_u$.

Let $\{e_i\}$ be an orthonormal basis for $S$.  First note 
that for any unit vector 
$v \in S$, the matrix of the projector $P_v$ on $v$ is obtained from 
\beq
\lbl{eqA80}
\lb e_j, P_v e_k \rb = \lb e_j, \lb v, e_k \rb v \rb
= \lb v, e_k \rb \lb e_j, v \rb.
\eeq

From the definition \re{eqA60}, for any fixed $j, k$ 
\begin{eqnarray}
\lbl{eqA90}
\lb e_j, (\trM P_u) e_k \rb &=& \sum_{\alpha} \lb e_j \tnr f_{\alpha} , 
P_u (e_k \tnr f_{\alpha}) \rb
\nonumber \\
&=&\sum_\alpha \lb u, e_k \tnr f_\alpha \rb 
\lb  e_j \tnr f_{\alpha}, u \rb \nonumber \\
&=&  
 \sum_{\alpha} \lb s_\alpha, e_k \rb \lb e_j, s_\alpha\rb 
\nonumber \\
&=& \sum_{\alpha} |s_{\alpha}|^2 
\lb \frac{s_{\alpha}}{|s_\alpha|},  e_k \rb  
\lb e_j, \frac{s_{\alpha}}{|s_\alpha|} \rb  
\nonumber \\
&=& \sum_{\alpha} |s_{\alpha}|^2 \lb e_j, P_{s_{\alpha}} e_k \rb
\end{eqnarray}
In summary,
\beq \lbl{eqA100}
\trM P_u = \trM P_{\sum_{\alpha} s_{\alpha} \tnr f_{\alpha}}
= \sum_{\alpha} |s_{\alpha}|^2 P_{s_{\alpha}} \q.
\eeq
(Recall that we are using the convention that $P_s = P_{s/|s|}$.)
This exhibits $\trM P_u$ as a convex linear combination of pure states 
$P_{s_{\alpha}}$.  (This makes it almost obvious that $\trM P_u$ is pure
if and only $u$ is a product state, a fact mentioned in the main text.) 

It is also of interest to calculate $ \trS P_u $
for the  $u = \sum_{\alpha} s_{\alpha} \tnr f_{\alpha}$ given by \re{eqA70}, 
with $|u| = 1$.  
Let $\{e_i\}$ 
and $\{f_j\}$ be orthonormal bases for $S$ and $M$, respectively.  Then
the matrix of $\trS P_u$ with respect to $\{ f_j \}$ is given by
\begin{eqnarray}
\lbl{eqA110}
\lb f_k, (\trS P_u) f_j \rb &=& \sum_\alpha \lb e_{\alpha} \tnr f_k ,
P_u (e_\alpha \tnr f_j) \rb \nonumber\\
&=& \sum_{\alpha} \lb e_{\alpha} \tnr f_k, \lb 
 \sum_{\beta} s_{\beta} \tnr f_{\beta} , e_{\alpha} \tnr f_j \rb 
\sum_{\gamma} s_{\gamma} \tnr f_{\gamma}
\rb \nonumber\\
&=&\sum_{\alpha} \lb s_j, e_{\alpha} \rb 
\lb e_{\alpha} , s_k \rb \nonumber \\
&=& \lb s_j, s_k \rb   
\q,
\end{eqnarray} 
where the last equality is Parseval's equality.

The statement and proof 
of equation \re{eqA110} assumed for simplicity that $|u|=1$, 
but of course the case of 
arbitrary nonzero $u$ can be immediately obtained by normalization. 
Since this result will be needed in Appendix 2 , we state it 
explicitly for the reader's convenience. Let $\{ f_i \}$ 
be an orthonormal basis for $M$.  Then for any nonzero vector
$u = \sum_{\alpha} s_{\alpha} \tnr f_{\alpha} \in S \tnr M$,
the matrix $(\lb f_k , (\trS P_u) f_j \rb ) $  of $\trS P_u$ is given by 
\beq
\lbl{eqA115} 
\mbox{for $u = \sum_{\alpha} s_{\alpha} \tnr f_{\alpha}$,} \q\q
\lb f_k , (\trS P_u) f_j \rb = 
\frac{ \lb s_j, s_k \rb}{|u|^2} \q. 
\eeq

\section{Appendix 2:  Issues in proving  weakness of the AAV-type  ``weak measurement''} 
Subsection 7.1 outlined a weak measurement protocol and gave 
a hand-waving motivation (equation \re{eq255}) why it it might be hoped 
to be weak in the sense of Definition \ref{def1}.  When examined in detail, 
the ``weakness'' issue turns out to be unexpectedly subtle. 

So far as I know, this issue has never been recognized in the physics 
literature. This subsection discusses it in detail.  

We shall be discussing the situation discussed in the 
main text 
in which the state space $S$ of primary interest is finite dimensional, but 
the meter space $M = L^2(\bR)$ is infinite dimensional. 
We switch to the notation $|| m || := [\,\int_\bR \, |m(q)|^2\, dq\,]^{1/2} $ 
for the $L^2$ Hilbert space norm of 
$m \in L^2(\bR)$, in order to reserve $|m(q)|$ for the absolute value 
of the function $m(\cdot) \in L^2(\bR)$.  We continue to use 
$|s| := \lb s,s \rb^{1/2}$ and $|u|$ for the norms of vectors $s \in S$
and $u \in S \tnr M$.

The notation will be as in the main text,  summarized  here for the
reader's convenience.  
Let $s \tnr m$ be a given product state in $S \tnr M$ with 
$|s| = 1 = ||m||,$ which will be 
fixed throughout the discussion. 
Let $H(\epsilon ) := \epsilon A \tnr P$, 
where $A$ is a given Hermitian operator on $S$, $P$ the  momentum operator on
$M = L^2(\bR)$, and $V$ a unitary operator on $S$ with $Vs = s$.

Recall that we 
apply $(V \tnr I) e^{-iH(\epsilon)}$ to $s \tnr m$ to obtain a state 
$(V \tnr I) e^{-iH(\epsilon)}(s\tnr m)$ 
in which the expectation 
\beq
\lb 
(V \tnr I) e^{-iH(\epsilon)}(s\tnr m),  
(I \tnr Q) (V \tnr I) e^{-iH(\epsilon)}(s\tnr m) \rb
\eeq
is to be measured, where $Q$ operates on $g \in L^2(\bR)$ by 
$(Qg)(q) := q g(q)$.   The hope is that this expectation divided by $\epsilon$
will approximate
$\lb s, As \rb$ for small $\epsilon$, and that the measurement will negligibly
change the starting state $s$ of $S$. 

At first glance, 	
	it seems plausible that for small $\epsilon$, the state  
	$(V \tnr I) e^{-iH(\epsilon)}(s\tnr m)$ would be close to $s \tnr m$, 
and that
	the corresponding states of $S$, namely  
	$\trM (V \tnr I) (e^{-iH(\epsilon)}(s\tnr m))$ and $\trM (s \tnr m) = s$ 
	would also be close.  
But the reader who attempts to rigorously justify this
	expectation will find that careful thought is necessary.  One 
	complication is that 
	the operation $\trM$ operates on an infinite dimensional space
	$S \tnr M$, so that the sense in which it is continuous has to be carefully
	considered.%
\footnote{
For example, for infinite dimensional $M$, 
	the mapping $ u\tnr v \goesto \trM P_{u \tnr v}$ 
	is not continuous with respect to the norm topologies, so 
	$e^{-iH(\epsilon)}(s\tnr m) \goesto s\tnr m $ as $\epsilon 
	\goesto 0$ does not immediately imply that 
$\trM P_{e^{-iH(\epsilon)(s \tnr m)}}  
\goesto \trM P_{s \tnr m} = P_s$.
}  
But one welcome simplification is that the definition of ``closeness'' 
	in $S$ (or in density matrices on $S$) 
	is not in issue:  all norms on a finite dimensional space are 
	equivalent, so we may define ``closeness'' in $S$ 
	by any convenient norm, and 
	reflect this in language which would otherwise be sloppy.  

{\em A priori}, it is not enough to show that 
	$\trM P_{(V \tnr I)e^{-iH(\epsilon)}(s\tnr m)} \goesto  P_s $ 
as $\epsilon \goesto 0$
	because of the possibility that  
	the measurement of $I \tnr Q$ may affect the (mixed) state  
	 $\trM P_{(V \tnr I)e^{-iH(\epsilon)}(s\tnr m)}$ of $S$.

Since $I \tnr Q$ has 
continuous spectrum, the precise manner in which the measurement of $Q$ 
affects the state
is not even defined by usual formulations of quantum mechanics.%
\footnote{If one tries to substitute ``Dirac delta functions'' for eigenvectors
of $Q$ one is led outside the Hilbert space $L^2(\bR)$.  For example, 
if one measures 
$Q$ and obtains the value 3.2, one cannot say that the subsequent state of the
system is $q \mapsto \delta(q - 3.2)$.  So far as the author knows, this 
problem has never been resolved in a complete and rigorous way.
}

It {\em would} be
defined if  $Q$ had pure point spectrum, so to carry out a rigorous analysis
we shall need to approximate $Q$ by an operator with pure point spectrum.  
To approximate $Q$ in norm within $\lambda > 0$ by such an operator $B$, 
define $B$ for $g \in L^2(\bR)$ by 
	\beq
	\lbl{eq 270}
	(Bg)(q) := \lambda [q/\lambda] g(q)
\q,
\eeq
where the bracket denotes the greatest integer function: for $x \in \bR, [x]$ 
denotes the greatest integer less than or equal to $x$.  The function
  $Bg$ need not be 
in $L^2(\bR)$, but restricting to the set of all $g$ for which $Bg$ {\em is}
in $L^2(\bR)$ makes $B$ Hermitian (we omit the details). 

Then $B$ has the 
set $\{k\lambda\}^{\infty}_{k = -\infty} $ of integer multiples of 
$\lambda$ as pure point spectrum.  
For each integer $k$, 
let $P_k$ denote the projector on the eigenspace of $B$ with eigenvalue 
$k \lambda $.
This eigenspace consists of all functions with support in 
$[k \lambda, (k+1) \lambda)$.   

A measurement of $B$ is the same as a projective
measurement with respect to the resolution of the identity 
$\{I\tnr P_k\}_{k=-\infty} ^\infty$.   
If the composite system is in normalized pure state $r \in S \tnr M$ 
before a measurement
of $B$,  then after the measurement, it will be in (unnormalized) 
pure state $(I \tnr P_k) r $ with probability $|(I \tnr P_k) r |^2$, 
i.e., in mixed state
$\sum_k |(I \tnr P_k) r|^2 P_{(I\tnr P_k) r}$.  
The state of $S$ will then be 
\beq
\lbl{eq10A2}
\sum_k |(I \tnr P_k) r|^2 \trM P_{(I\tnr P_k) r}
\q.
\eeq

Let $\{a_j\}_{j=1}^n$ be an orthonormal basis for $S$ of eigenvectors $a_j$
of $A$: $ A a_j = \alpha_j a_j$. 
For our given $s \in S$, let $s = \sum_j \sigma_j a_j$ be its expansion
as a linear combination of these eigenvectors.
We are interested in the case 
\beq
\lbl{eq20A2}
r := (V \tnr I) e^{-i H(\epsilon)} (s \tnr m)
=\sum_j \sigma_j Va_j\tnr m_{\epsilon \alpha_j}
=\sum_j  Va_j\tnr \sigma_j m_{\epsilon \alpha_j}, 
\eeq
where $m_\beta$ denotes the translate of $m$ by $\beta$: 
$m_{\beta}(q) := m(q - \beta)$ for all $q \in \bR$.
By formula \re{eqA115}  of Appendix 1 (with the roles of $S$ and $M$ reversed),
the $i,j$ matrix element of $\trM P_{(I \tnr P_k) r}$  with respect to the 
orthonormal basis $\{Va_i\}$ is: 
\begin{eqnarray}
\lbl{eq30A2}
\lefteqn{\lb Va_i , \trM P_{(I \tnr P_k) r} Va_j \rb} \nonumber \\ 
&=&    
\frac{\lb P_k m_{\epsilon \alpha_j}, P_k m_{\epsilon \alpha_i} \rb
\sigma^*_j \sigma_i
} 
{ |(I \tnr P_k) r|^2}. 
\end{eqnarray}
Substituting in \re{eq10A2} gives the state of $S$ after the projective
measurement, 
expressed as a density matrix with respect to the orthonormal basis 
$\{Va_j\}$, as
\beq
\lbl{eq40A2}
 \sum_k 
\lb P_k m_{\epsilon \alpha_j}, P_k m_{\epsilon \alpha_i} \rb
\sigma^*_j \sigma_i \q.
\eeq
Our goal is to show that the limit as $\epsilon \goesto 0$ of expression 
\re{eq40A2} is $P_s$ (which is $s$ expressed as a density matrix).  
This is immediate if it is legitimate to interchange this limit with the
summation in \re{eq40A2} because
\begin{eqnarray}
\lbl{eq50A2} 
\lefteqn{ \sum_k 
\lim_{\epsilon \goesto 0} 
\lb P_k m_{\epsilon \alpha_j}, P_k m_{\epsilon \alpha_i} \rb
\sigma^*_j \sigma_i }\nonumber\\
&=& \sum_k  \lb P_k m, P_k m \rb \sigma^*_j \sigma_i  \nonumber\\
&=&  \sum_k ||P_k m||^2 \sigma^*_j \sigma_i =  \sigma^*_j \sigma_i , 
\end{eqnarray}
since $\{P_k\}$ is a resolution of the identity and $||m|| = 1$.
The operator on $S$ whose matrix  with respect to the basis 
$\{Va_j \}$ is given by \re{eq50A2} as $(\sigma^*_j \sigma_i)$ is the projector
onto the vector 
$$
\sum_j \sigma_j Va_j = V \sum_j \sigma_j a_j = Vs = s 
\q.
$$  
Thus we have established weakness of the measurement of $B$ 
if the meter state $m$ satisfies
\beq
\lbl{eq60A2}
\lim_{\epsilon \goesto 0} \sum_k 
\lb P_k m_{\epsilon \alpha_j}, P_k m_{\epsilon \alpha_i} \rb
= \sum_k
\lim_{\epsilon \goesto 0} 
\lb P_k m_{\epsilon \alpha_j}, P_k m_{\epsilon \alpha_i} \rb
\q.
\eeq

It is legitimate to interchange limits 
in an expression like \re{eq60A2} if all of the indicated limits exist and if at least
one of the inner limits is uniform.%
\footnote{In case the reader is puzzled that there only appears
to be one limit in \re{eq60A2}, note that the infinite sum implictly includes
a limit.}
A simple hypothesis which assures this is that
$m$ have compact support.  In that case, the sum on the left side of \re{eq60A2}  
is actually finite with a maximum number of nonzero terms independent of 
$\epsilon$ for small $\epsilon$, so the situation is trivial.  
Since physical ``meters'' have a bounded range of possible readings, this
hypothesis is physically reasonable.

Of course, many 
other conditions will also suffice.   For those who like to work with 
Gaussians, if $m$ is a Gaussian (or more generally,
if $m(q)$ decays sufficiently rapidly as $ q \goesto \pm \infty$), uniform 
convergence of the sum follows from  brute force upper bounds on $|m(q)|$
for large $q$.  

As for existence of the limits, 
the limits in the right-hand expression in \re{eq60A2}
obviously exist.  If the left-hand limits are in question, one could pass
to a sequence $\{\epsilon_l\}^\infty_{l=1}$ with $\epsilon_l \goesto 0$ for which 
$\lim_{l \goesto \infty}   
\sum_k \lb P_k m_{\epsilon_l \alpha_j}, P_k m_{\epsilon_l \alpha_i} \rb
$
does exist.  (Routine compactness arguments assure the existence of such
a subsequence.)  Then the uniformity shows that all the double limits are equal
(and  that the left-hand limit in \re{eq60A2} actually does exist).



\end{document}